\documentclass[
 reprint,
 amsmath,amssymb,
 aps,
]{revtex4-2}

\usepackage{graphicx}% Include figure files
\usepackage{float}
\usepackage{dcolumn}% Align table columns on decimal point
\usepackage{amsfonts} 
\usepackage{dsfont}
\usepackage{amssymb}
\usepackage{svg}
\usepackage{amsmath}
\usepackage{hyperref}
\usepackage{subcaption}
\usepackage{slashed}
\usepackage[absolute,overlay]{textpos} % For absolute positioning
\usepackage{placeins}
\usepackage{dsfont}
\begin{document}

\preprint{APS/123-QED}

\title{Quantum Estimation in QED Scattering}

\author{Preslav Asenov}
\email{preslav.asenov.20@ucl.ac.uk}
\author{WenHan Zhang}
\author{Alessio Serafini}
\email{serale@theory.phys.ucl.ac.uk}

\affiliation{Department of Physics and Astronomy, University College London,
 Gower Street, London WC1E 6BT, United Kingdom}

\date{\today}% It is always \today, today,

\begin{abstract}
We tackle the issue of estimating dynamical parameters in quantum electrodynamics. We numerically compute the quantum Fisher information matrix (QFIM) of physical parameters in electron-muon and Compton scattering at tree level. In particular, we consider the estimation of centre-of-mass three-momentum magnitude and polar scattering angle through measurements on the internal degrees of freedom (helicity or polarisation) of the scattered particles. Computations are carried out for pure and maximally mixed initial states. The QFIM values are
then used to compute the quantum Cramér-Rao lower bounds on the estimations at hand. Further, we compare such ultimate bounds to the classical Fisher information of local polarisation or helicity degrees of freedom.

\end{abstract}

\maketitle

\renewcommand{\arraystretch}{1.5}
%\tableofcontents

\section{Introduction}
Making inferences based on experimental measurements is a fundamental step of the scientific method \cite{platt1964strong}. One example of such an inference is the estimation of one or multiple parameters on which a measurement may depend \cite{westcott1960theparameter}. This is referred to as the parameter estimation problem and is approached using the methods of estimation theory \cite{paris2009quantum}. In the context of quantum measurements, it is the framework of quantum metrology, or quantum estimation theory (QET), that offers methods to establish the precision limits constraining quantum measurements, as well as, in principle, schemes that achieve these limits and potentially surpass the performance of classical systems \cite{giovannetti2011advances}. A key concept in QET is the quantum Fisher information (QFI) of a quantum state, calculated with respect to the parameters to be estimated \cite{paris2009quantum}. The QFI allows one to place a fundamental precision bound on parameter estimation, known as the Cram\'{e}r-Rao lower bound.

QFI calculations have been applied in a variety of contexts, including condensed matter physics \cite{dellanna2023quantum, mazza2024quantumfisherinformationstrange}, cosmology \cite{gomez2021quantum, chen2024quantum}, and optics \cite{zhang2012unbounded, tsang2019resolving, hradil2019quantum}. The QFI has also been used to show a connection between classical fluid dynamics and relativistic quantum mechanics \cite{yahalom2023afluid}.

The growing interest in the connection between quantum information and high-energy physics (HEP) has also led to investigations of the QFI in HEP. Quantum information metrics \cite{blau2001instantons, miyaji2015distance, tsuchiya2022geometrical, erdmenger2022complexity} and the QFI in particular \cite{ bak2016information, trivella2017holographic, lashkari2016canonical, banerjee2018connecting, erdmenger2020information, dimov2021holographic}, have been studied in the context of the anti-de Sitter/conformal field theory (AdS/CFT) correspondence too. The QFI of an Unruh-DeWitt detector has also been investigated in Ref. \cite{feng2022quantum}. However, there is no known study examining the QFI of excitations of the quantum field in the context of HEP scattering. This is the central issue addressed by the present paper, as we consider QFI in QED scattering at the tree level. This is a development of previous investigations considering the quantum informational properties of QED scattering. In this specific regard, Refs. \cite{cerveralierta2017maximal, cerveralierta2019maximalentanglementapplicationsquantum} investigated entanglement between the helicity degrees of freedom in tree-level QED, in particular focusing on the conditions for maximal entanglement, whilst Ref. \cite{fedida2023treelevel} investigated such quantum correlations over a wider range of scattering energies, for pure and mixed scattering states. 

The tree-level QED scattering processes we shall consider in this paper are electron-muon and Compton scattering, which will be prototypical of QED processes. 

The specific estimation we will be considering is better explained by employing the services of Alice and Bob, the prototypical actors of experiments in the quantum information tradition. In our scenario, Alice prepares two initial particles with known centre of mass energy and lets them scatter off each other. Hence, she selects the outgoing scattering angle with infinite precision, say through an infinitely narrow collimator. The two-particle quantum state she thus prepares and sends off (in opposite directions) is characterised by the parametres $p$, the conserved total three-momentum magnitude in the centre of mass frame, and $\theta$, the selected scattering angle. Bob receives this state, but has no knowledge of the two parameters. In order to reconstruct them, he can only access internal degrees of freedom (polarisation for photons, helicity for massive particles). We assume such an access to be global, that is, Bob can perform global measurements on the internal two-qubit state he receives from Alice (but then also consider a local estimation scheme). We therefore consider a multi-parameter quantum estimation of the external degrees of freedom of the scattering process, namely the three-momentum magnitude and the scattering angle, based on such internal degrees of freedom quantum measurements. To this aim, we calculate the quantum Fisher information matrix (QFIM) of the output two-qubit state after scattering, whose inverse provides a lower bound on the covariance matrix associated with the estimation of the external parameters considered, by virtue of the quantum Cramér--Rao bound. Thus, we place ultimate, rigorous bounds on the estimation procedure. Besides, we also provide a specific example of estimation, based on diagonal helicity or polarisation measurements, and contrast it with the general bounds.

Albeit the scenario we just described is not, per se, directly relevant in practice \footnote{Notice however that, although, one could argue, Bob could obtain direct information on the outgoing momenta through a calorimeter and detector arrays (where time-coincidences would reveal the propagation direction of particle pairs, as customary, for instance, in PET scans), our method would reveal additional information on the original propagation direction of the incoming scattering pair.}, it does lend an exact operational meaning to the bounds we will derive. On a fundamental level, our investigation addresses the hitherto-untackled question of how much information about the external degrees of freedom can be gleaned by accessing the internal degrees of freedom of a quantum field. As we shall discuss in our conclusions, our approach may be further developed and extended to improve estimation in cases with direct practical impact.

\section{Setting and Preliminaries}
In order to provide the relevant set-up for this paper, Section \ref{Fisher_information} introduces the problem of multi-parameter quantum estimation and the concepts of classical and quantum Fisher information. Section \ref{QE_in_QED_scattering} states the estimation process that we consider. Section \ref{QED_scattering} provides the derivation of the post-scattering, four-momentum-filtered quantum state density matrix, whose QFIM is then calculated. Finally, Section \ref{sec:Optimal_measurement_of_pure_states} provides the optimal positive operator-valued measure (POVM) for the measurement of pure quantum states.

\subsection{Fisher Information}\label{Fisher_information}
\subsubsection{Classical Fisher Information}
In estimation theory we consider a vector of physically observable random variables $\textbf{X} = ( X_1, X_2, \dots, X_n)$ \cite{devore2011probability} and define $\textbf{x} = ( x_1, x_2, \dots, x_n)$ to be a vector of values that $\textbf{X}$ take after a single measurement is performed. This measurement process can then be repeated, giving us a set of different $\textbf{x}$ values. The probability that $\textbf{X} = \textbf{x}$ for a given measurement is described by the joint probability density function (PDF) $p(\textbf{x}) = p(\textbf{x}|\boldsymbol{\sigma})$, which depends on the vector of parameters $\boldsymbol{\sigma} = ( \sigma_1, \sigma_2, \dots, \sigma_m)$. Within this framework, the task of accurately estimating $\boldsymbol{\sigma}$, given a set of $\textbf{x}$ measurements, is rephrased as the task of finding a suitable estimator $\hat{\boldsymbol{\sigma}}(\textbf{x})$, which is close to the true $\boldsymbol{\sigma}$ \cite{helstrom1969quantum}. $\hat{\boldsymbol{\sigma}}(\textbf{x})$ is defined as a mapping 
\begin{equation}
    \hat{\boldsymbol{\sigma}}: \boldsymbol{\mathcal{X}} \rightarrow \boldsymbol{\Sigma},
\end{equation}
where $\boldsymbol{\mathcal{X}}$ is the space of observations $\textbf{x}$, and $\boldsymbol{\Sigma}$ is the space  of parameter values $\boldsymbol{\sigma}$ \cite{zielinski1997theory}.\\

If the measurement performed on $\textbf{X}$ is classical, then the study of $\hat{\boldsymbol{\sigma}}(\textbf{x})$ is subject to classical estimation theory (CET). In this paper we consider the properties of unbiased estimators, which are defined to have an expectation value equal to the true parameter value \cite{devore2011probability}:
\begin{equation} \label{unbiased_estimator}
    \text{E} \left[ \hat{\boldsymbol{\sigma}}(\textbf{x}) \right] = \boldsymbol{\sigma}.
\end{equation}
Without making any further assumptions about the properties of $\hat{\boldsymbol{\sigma}}(\textbf{x})$, we are interested in lower bounding the precision of the estimation $\hat{\boldsymbol{\sigma}}(\textbf{x})$ performs. For a classical measurement, this precision lower bound is placed on the variance of $\hat{\boldsymbol{\sigma}}(\textbf{x})$ (i.e. $\text{Var} [ \hat{\boldsymbol{\sigma}}(\textbf{x}) ]$) and is defined as the classical  Cram\'{e}r-Rao lower bound (CCRLB). In order to define the CCRLB we introduce the classical Fisher information matrix $\textbf{I}^{(C)}_{\boldsymbol{\sigma},\textbf{x}}$ (CFIM). The following review of the CFIM is adapted from \cite{ly2017tutorial}. The CFIM is a positive semi-definite, symmetric $m \times m$ matrix, with its $(i,j)$ element given by
\begin{equation}
    \left( \textbf{I}^{(C)}_{\boldsymbol{\sigma},\textbf{x}} \right)_{i,j} =
    \text{Cov} \left[L^{(C)}_{\sigma_i} , L^{(C)}_{\sigma_j} \right].
\end{equation}
Here $L^{(C)}_{\sigma_i} = \frac{\partial\ \text{log}\ p(\textbf{x}|\boldsymbol{\sigma})}{\partial \sigma_i}$ is defined as the score function, with its derivative evaluated at the true value of $\sigma_i$ \cite{pickles1985introduction,kay1993fundamentals}, while the covariance $\text{Cov} \left[ X,Y \right]$ of two random variables $X$ and $Y$ is defined as
\begin{align} \label{covariance_definition}
    \text{Cov} \left[ X,Y \right] = \text{E}\left[ \left( X - \text{E}\left[ X \right] \right) \left( Y - \text{E}\left[ Y \right] \right) \right].
\end{align}
Accordingly, $\left[ X,X \right]=\text{Var} \left[X\right]$. 
In the single-parameter estimation case, where $\boldsymbol{\sigma} = \sigma$, the CFIM $\textbf{I}^{(C)}_{\boldsymbol{\sigma},\textbf{x}}$  reduces to the classical Fisher information (CFI) $\text{I}^{(C)}_{\sigma,\textbf{x}}$ given by
\begin{align} \label{fisher_info_definition1}
    \text{I}^{(C)}_{\sigma,\textbf{x}} &= \text{Var} \left[ L^{(C)}_{\sigma}(\textbf{x}) \right] \nonumber \\[5pt]
    & = \text{E} \left[ \left( L^{(C)}_{\sigma}(\textbf{x}) \right)^2 \right],
\end{align}
where $L^{(C)}_{\sigma}(\textbf{x}) = \frac{\partial\, \text{log}\, p(\textbf{x}|\sigma)}{\partial \sigma}$.
%Here we assume that the PDF $p(\textbf{x}|\boldsymbol{\sigma})$ is such that $L^{(C)}_{\sigma}(\textbf{x})$ satisfies the regularity condition
%\begin{equation} \label{score_regularity}
%    \text{E} \left[ L^{(C)}_{\sigma}(\textbf{x}) \right] = 0,
%\end{equation}
%where $E[\cdot]$ is the expectation value taken with respect to $p(\textbf{x}|\boldsymbol{\sigma})$ \cite{kay1993fundamentals}. Under this condition $\text{I}^{(C)}_{\sigma,\textbf{x}}$ can be written as 
%\begin{equation} \label{fisher_info_definition2}
%    \text{I}^{(C)}_{\sigma,\textbf{x}} = \text{E} \left[ \left( L^{(C)}_{\sigma}(\textbf{x}) \right)^2 \right].
%\end{equation}
In order to intuitively understand the motivation behind the definition of $\text{I}^{(C)}_{\sigma,\textbf{x}}$ as a quantifier of information about the value of $\sigma$, it is useful to consider a scenario where we collect $N$ different measurements $\textbf{x}^{(i)}$ for $i \in \{1, \dots ,N\}$ and then plot $N$ different $\text{log}p(\textbf{x}^{(i)}|\sigma)$-$\sigma$ curves, where $\text{log} p(\textbf{x}^{(i)}|\sigma)$ is defined as the log-likelihood \cite{wassermann2004allof} of the respective curve. In these plots, the PDF $p(\textbf{x}^{(i)}|\sigma)$ tells us the probability of the parameter $\sigma$ being equal to any given value, for a fixed measurement result $\textbf{x}^{(i)}$. The next step is to compute the score function $L^{(C)}_{\sigma}(\textbf{x})$ of each curve. $L^{(C)}_{\sigma}(\textbf{x})$ can be thought of as a piece of information about the shape of the curve, or as an ``instruction'' about where to find its local minimum. If the $N$ score functions $L^{(C)}_{\sigma}(\textbf{x})$ have a large variance of values around a given $\sigma$, then we have a distribution of sharp $\text{log}p(\textbf{x}^{(i)}|\sigma)$-$\sigma$ curves, meaning that $p(\textbf{x}|\sigma)$ is sensitive to small changes in $\sigma$, hence $p(\textbf{x}|\sigma)$ contains substantial information about $\sigma$. This corresponds to a large Fisher information value $\text{I}^{(C)}_{\sigma,\textbf{x}}$ around a given $\sigma$. Conversely, a small variance of score functions at a given $\sigma$ corresponds to a large spread of broad curves, i.e. $p(\textbf{x}|\sigma)$ contains less information about $\sigma$, which is quantified by a smaller $\text{I}^{(C)}_{\sigma,\textbf{x}}$.\\

Having motivated the definition of $\text{I}^{(C)}_{\sigma,\textbf{x}}$, we now derive the CCRLB in a single-parameter estimation scenario. The CCRLB is ultimately of a geometric origin, as it arises from the Cauchy–Schwarz (CS) inequality. The CS inequality for two real-valued random variables $X$ and $Y$, with an inner product defined by $\langle X, Y \rangle := \text{E} \left[ XY \right]$, is given by
$
    \text{E} \left[XY \right]^2 \leq  \text{E} \left[X^2 \right] \text{E} \left[Y^2 \right]
$, which leads to
$
    \text{Cov}\left[X,Y \right]^2 \leq  \text{Var} \left[X \right] \text{Var} \left[Y \right].
$
Upon substituting $X \rightarrow \hat{{\sigma}}(\textbf{x})$ and $Y \rightarrow L^{(C)}_{\sigma}(\textbf{x})$, the covariance inequality yields
\begin{align} \label{cramer_rao_part1}
    \text{Cov}\left[\hat{{\sigma}}(\textbf{x}),L^{(C)}_{\sigma} (\textbf{x}) \right]^2 &\leq \text{Var} \left[\hat{{\sigma}}(\textbf{x}) \right]  \text{I}_{\sigma, \textbf{x}}^{(C)}.
\end{align}
%Using the definition of covariance (\ref{covariance_definition}) again, we have
%\begin{equation} \label{estimator_score_covariance_1}
%   \text{Cov}\left[\hat{\sigma},L^{(C)}_{\sigma}  \right] = 
%    \text{E}\left[ \left( \hat{{\sigma}} - \text{E}\left[ \hat{{\sigma}} \right] \right) \left( L^{(C)}_{\sigma}  - \text{E}\left[ L^{(C)}_{\sigma} \right] \right) \right].
%\end{equation}
Using the property $ \text{E} \left[ L^{(C)}_{\sigma}(\textbf{x}) \right] = 0$ and the expectation value of an unbiased estimator (\ref{unbiased_estimator}) leads to
\begin{align}
    \text{Cov}\left[\hat{{\sigma}},L^{(C)}_{\sigma} \right] &= 
    \text{E}\left[\hat{{\sigma}} L^{(C)}_{\sigma} \right]
    = \text{E}\left[ \hat{\sigma} \cdot \frac{\partial p}{\partial \sigma} \frac1p \right].
\end{align}
In the case of a continuous random variable $\textbf{x}$, this yields
\begin{align}\label{estimator_score_covariance_2}
    \text{Cov}\left[\hat{{\sigma}},L^{(C)}_{\sigma} \right] &= 
    \int \hat{\sigma} (\textbf{x}) \cdot \frac{\partial p(\textbf{x}|\sigma)}{\partial \sigma} d\textbf{x} = 1 ,
\end{align}
where we allowed the interchange of derivative and integral and then applied the unbiased estimator condition (\ref{unbiased_estimator}). Hence, by substituting (\ref{estimator_score_covariance_2}) into (\ref{cramer_rao_part1}) one obtains the expression of the CCRLB \cite{kay1993fundamentals}:
\begin{equation}
    \text{Var}\left[\hat{\boldsymbol{\sigma}}(\textbf{x})\right] \geq \frac{1}{\text{I}^{(C)}_{\sigma,\textbf{x}}}.
\end{equation}
The Fisher information is by definition additive over repeated, independent samplings of the distribution, so that for $N$ such samplings one has the CCRLB $\text{Var}\left[\hat{\boldsymbol{\sigma}}(\textbf{x})\right] \geq {1}/\left({N\text{I}^{(C)}_{\sigma,\textbf{x}}}\right)$.
The Cram\'er-Rao bound sets the ultimate limit to the precision achievable on the estimation of a parameter (at least in terms of its estimation variance).

The multi-parameter case can be treated along the same lines, although there the CCRLB takes the form of a matrix inequality relating the inverse of the $n\times n$ matrix $\textbf{I}^{(C)}_{\boldsymbol{\sigma},\textbf{x}}$ to the covariance matrix $\text{Cov}[\hat{\boldsymbol{\sigma}}(\textbf{x})]$ of the multi-parameter estimator $\hat{\boldsymbol{\sigma}}(\textbf{x})$, as
\cite{kay1993fundamentals}:
\begin{equation}
   \text{Cov}\left[\hat{\boldsymbol{\sigma}}(\textbf{x})\right] \geq \left[\textbf{I}^{(C)}_{\boldsymbol{\sigma},\textbf{x}}\right]^{-1}.
\end{equation}
It is very instructive to consider a two-parameter instance of this estimation:
\begin{equation} \label{eq:2_by_2_qcrlb}
\begin{pmatrix}
\text{Var} [\hat{x} ] & \text{Cov} [\hat{x},  \hat{y} ]\\[5pt]
\text{Cov} [ \hat{x} , \hat{y}] & \text{Var} [ \hat{y} ] 
\end{pmatrix} \geq \frac{1}{\left( \textbf{I}_{xx}\textbf{I}_{yy} - \textbf{I}_{xy}^2 \right)}
\begin{pmatrix}
\textbf{I}_{yy} & -\textbf{I}_{xy}\\[5pt]
-\textbf{I}_{xy} & \textbf{I}_{xx} 
\end{pmatrix}, 
\end{equation}
where we write $\textbf{I}^{(C)}_{\boldsymbol{\sigma}, \textbf{x}} = \textbf{I}_{\boldsymbol{\sigma}}$ for simplicity and we set $\sigma_1=x$ and $\sigma_2=y$.
First off, notice that, as one should expect, in the absence of correlations between the two variables (i.e., for $\text{Cov} [ \hat{x} , \hat{y}]=0$), one recovers two independent single-parameter CCRLB's in terms of the respective Fisher informations $I_{xx}$ and $I_{yy}$. Further, notice that the bound is generally worse than in the single-parameter case -- i.e., the fundamental lower bounds on the individual variances, on the matrix's main diagonal, are larger than the single-parameter bounds. The single-parameter bounds are also recovered if the Fisher information on the other parameter diverges -- i.e., letting $I_{yy}\rightarrow\infty$ yields 
\begin{equation} \label{eq:var_x_fixed_y}
    \text{Var} [\hat{x} ]_{\text{fixed}\, y} \ge \frac{1}{I_{xx}}.
\end{equation}

\subsubsection{Quantum Fisher Information}
Let us now move on to {\em quantum} estimation theory, by considering the case of a quantum state $\rho (\sigma)$, rather than a classical probability distribution, depending on a parameter $\sigma$. The key realisation here is that, once a positive operator valued measure (POVM), i.e., a specific quantum measurement, is set, the Born rule extracts a classical probability distribution from a quantum state, and one can apply the classical analysis presented above, whereby the variance of a parameter $\sigma$ is minimised by the (achievable) CCRLB. The optimisation of the CCRLB over all possible POVMs then gives the quantum Cram\'er-Rao lower bound (QCRLB), in terms of the quantum Fisher information. The latter is therefore just the classical Fisher information optimised over all possible POVMs.

 The following description of QET is adapted from \cite{paris2009quantum}. Again, let us treat first the single-parameter estimation case. In QET, we describe our system by the density operator  rather than the PDF $p(\textbf{x}|\sigma)$ \cite{helstrom1969quantum}. The two are related by the Born rule
    $p(\textbf{x}|\sigma) = \text{Tr} \left[ \Pi (\textbf{x}) \rho (\sigma) \right]$,
where $\Pi (\textbf{x})$ are POVM elements satisfying $\int \Pi (\textbf{x}) d\textbf{x} = {\mathds{1}}$. 

In QET, the (classical) score function $L^{(C)}_{\sigma}(\textbf{x})$ is generalised to the symmetric logarithmic derivative (SLD) quantum operator $L_{\sigma}$. Since, in the classical case, $L^{(C)}_{\sigma}(\textbf{x}) p(\textbf{x}|\sigma) = \partial_\sigma{p(\textbf{x}|\sigma)}$, 
in QET one sets 
\begin{equation}\label{sld_definition}
    \frac{1}{2} \left( L_{\sigma} \rho (\sigma) + \rho (\sigma)L_{\sigma} \right) = \partial_{\sigma} \rho (\sigma).
\end{equation}
Then, the quantum Fisher information (QFI) of state $\rho (\sigma)$ with respect to the parameter $\sigma$ is defined by \begin{equation}\label{quantum_fisher_info_definition1}
\text{I}(\sigma) = \text{Tr} \left[ \rho (\sigma) \left( L_{\sigma} \right)^2 \right],
\end{equation}
which is the quantum generalisation of the classical Fisher information expression in Eq. (\ref{fisher_info_definition1}), in the sense that it maximises the classical Fisher information over all POVMs. For $N$ independent quantum measurements of $\textbf{X}$ on $\rho (\sigma)$, the variance of any estimator $\hat{\boldsymbol{\sigma}}$ is in fact constrained by the quantum Cram\'{e}r-Rao lower bound (QCRLB):
\begin{equation}\label{eq:single_var_qcrlb}
    \text{Var}\left[\hat{\sigma}(\textbf{x})\right] \geq \frac{1}{N \text{I}_{\sigma}}.
\end{equation}

Analogously to its classical counterpart, in the multi-parameter estimation case, the QFI generalises to the quantum Fisher information matrix (QFIM) $\textbf{I}(\sigma)$. For $m$ parameters $\boldsymbol{\sigma} = ( \sigma_1, \sigma_2, \dots, \sigma_m)$ the QFIM is a positive semi-definite, symmetric $m \times m$ matrix, with its $(i,j)$ element given by
\begin{equation}
    \left( \textbf{I}_{\boldsymbol{\sigma}} \right)_{i,j} = \text{Tr} \left[ \rho (\boldsymbol{\sigma}) \frac{L_{\sigma_i} L_{\sigma_j} + L_{\sigma_j} L_{\sigma_i}}{2} \right].
\end{equation}
This expression is equivalent to \cite{liu2019quantum}
\begin{equation} \label{eq:Fisher_multiparam_mixed}
    \left( \textbf{I}_{\boldsymbol{\sigma}} \right)_{i,j} = 2 \sum_{k,l}^{d-1} \frac{\text{Re}\left[ \langle k| \partial_i  \rho (\boldsymbol{\sigma}) |l \rangle \langle l| \partial_j  \rho (\boldsymbol{\sigma}) |k \rangle \right]}{\lambda_k + \lambda_l},
\end{equation}
where $\partial_j  \rho (\boldsymbol{\sigma}) = \frac{\partial \rho (\boldsymbol{\sigma})}{\partial \sigma_i}$ and $\rho (\boldsymbol{\sigma}) = \sum_{k}^{d-1} \lambda_k(\boldsymbol{\sigma}) |k\rangle \langle k|$ is the eigendecomposition of the density matrix $\rho (\boldsymbol{\sigma})$. Here $\{|k\rangle\}$ and $\{ 
\lambda_k \}$ are, respectively, the sets of eigenstates and eigenvalues of $\rho (\boldsymbol{\sigma})$. For a pure state $\rho (\boldsymbol{\sigma}) = \rho^2 (\boldsymbol{\sigma}) = |\psi \rangle \langle \psi|$ Eq. (\ref{eq:Fisher_multiparam_mixed}) reduces to \cite{liu2019quantum}
\begin{equation} \label{eq:Fisher_multiparam_single}
    \left( \textbf{I}_{\boldsymbol{\sigma}} \right)_{i,j} = 4\text{Re}\left[ \langle \partial_i \psi |\partial_j \psi \rangle - \langle \partial_i \psi| \psi \rangle \langle \psi | \partial_j \psi \rangle \right] ,
\end{equation}
where $|\partial_j \psi \rangle = \frac{\partial | \psi \rangle}{\partial \sigma_j}$. In the multi-parameter estimation scenario we consider here, the QCRLB bounds the covariance matrix $\text{Cov} \left[ \hat{\boldsymbol{\sigma}}(\textbf{x}) \right]$ as:
\begin{equation}\label{eq:qcrlb_general}
    \text{Cov} \left[ \hat{\boldsymbol{\sigma}}(\textbf{x}) \right] \geq \frac{1}{N}{\textbf{I}_{\boldsymbol{\sigma}}}^{-1},
\end{equation}
where ${\textbf{I}_{\boldsymbol{\sigma}}}^{-1}$ is the inverse of the QFIM.

It is worthwhile adding that one of the (not necessarily unique) POVMs that achieves the QCRLB is that whose elements are single-rank projectors on the eigenvalues of the SLD operator. In principle, this provides a systematic method to identify an optimal measurement (although it may well be impossible to implement in practice).

Notice that the QFIM acts as a Riemannian metric on the space of quantum states, equivalent to the Bures distance between quantum states over the manifold spanned by the parameters under consideration (except for values at which the state's rank varies \cite{safranek17}). So, by computing the QFIM, we explore a property of the space of quantum states with respect to parameters encoded through the particular quantum field dynamics we consider.

\subsection{QED scattering and filtering}\label{QED_scattering}
Here, we present the steps required to prepare and describe the state of a quantum field whose external degrees of freedom are to be estimated based on a measurement of its internal degrees of freedom. This state preparation process consists of QED scattering and followed by a momentum measurement. 

We consider the scattering of two asymptotically free particles
%The unitary evolution of state $\rho_{t_0}$ at time $t_0$ to state $\rho_{t}$ at time $t$ is modeled by the application of the time-evolution operator $U(t, t_0)$:
%\begin{equation}
%    \rho_{t} = U(t, t_0) \rho_{t_0} U^{\dagger}(t, t_0).
%\end{equation}
%In the asymptotic limit, where $t \rightarrow \infty$ and $t_0 \rightarrow -\infty$, the time-evolution equation becomes
%\begin{equation}
%    \rho_{+\infty} = S \rho_{-\infty} S^{\dagger},
%\end{equation}
%where $S$ is the S-matrix operator, defined by
%\begin{equation}
%    S = \lim_{t \rightarrow + \infty}\lim_{t_0 \rightarrow -\infty} U(t,t_0).
%\end{equation}
for a two-particle initial state $\rho_{-\infty}$ with given centre-of-mass momentum magnitude and maximally mixed in the internal (spin) degrees of freedom:
\begin{equation} \label{eq:incoming_state_general}
    \rho_{-\infty} = \sum_{\lambda} \frac14 |{\bf p},\lambda\rangle \langle {\bf p},\lambda|.
\end{equation}
Here we use the short-hand $|{\bf p},\lambda\rangle = a^{\dag}_{{\bf p},\lambda_1}b^{\dag}_{-{\bf p},\lambda_2}|0\rangle$, where $a$ and $b$ are (bosonic or fermionic) field operators for three-momentum ${\bf p}$, while $\lambda \equiv (\lambda_1,\lambda_2) \in \{ LL, LR, RL, RR\}$. $L$ labels a left-handed helicity eigenstate for a fermion or a left-handed circular polarisation state for a photon, while $R$ labels right-handedness in a similar way. 

As already stated, we will consider a scenario where an observer measures the momentum $\bf q$ of a scattered particle (which, clearly, also fixes the momentum of the other one in the centre of mass frame) on the final, scattered state $\rho_{+\infty}$, by applying the POVM element $\Pi_{\bf q} = \sum_{\eta} |{\bf q},\eta \rangle \langle {\bf q}, \eta |$. This idealised (infinitely sharp in momentum) filtering process projects $\rho_{+\infty}$ onto a two-qubit state of the internal degrees of freedom that depends on the initial and final momenta ${\bf p}$ and ${\bf q}$. Our aim is to consider the quantum estimation of these dynamical parameters based on measurements on the internal degrees of freedom.

If $S$ is the QED S-matrix operator, i.e., the unitary relating the initial state to the final state after scattering, the resulting state after the momentum filtering process is
\begin{align} \label{filtered_state1}
    \rho_{\text{out}} &= \frac{ \Pi_{\bf q} S\rho_{-\infty}S^{\dag} \Pi_{\bf q}}{ \text{Tr}[ \Pi_{\bf q} S\rho_{-\infty}S^{\dag} \Pi_{\bf q} ] }\nonumber\\[1ex]
    &= \frac{\sum_{\lambda, \eta, \eta'} |{\bf q}, \eta\rangle \langle {\bf q}, \eta|S|{\bf p}, \lambda\rangle \langle {\bf p}, \lambda|S^{\dagger}|{\bf q}, \eta'\rangle \langle {\bf q}, \eta'| }{ \sum_{\lambda, \nu}  \langle {\bf q}, \nu|S|{\bf p}, \lambda\rangle \langle {\bf p}, \lambda|S^{\dagger}|{\bf q}, \nu\rangle}\nonumber\\[1ex]
    &= \sum_{\eta, \eta'} \rho_{\eta, \eta'} |{\bf q}, \eta \rangle \langle {\bf q}, \eta'|,
\end{align}
which can be represented by a $4 \times 4$ matrix in the helicity or polarisation basis. Eq. (\ref{filtered_state1}) shows that the entries of the two-qubit density matrix $\rho_{\eta, \eta'}$ can be written in terms of the S-matrix elements $\mathcal{M}_{{\bf p}\lambda \rightarrow {\bf q}\eta}$ that are evaluated through Feynman diagrams in the standard perturbative expansion, as per
\begin{equation}\label{eq:out_state_mat_elem}
\rho_{\eta, \eta'} = \frac{ \sum_{\lambda} \mathcal{M}_{{\bf p}\lambda \rightarrow {\bf q}\eta} \mathcal{M}^{*}_{{\bf p}\lambda \rightarrow {\bf q}\eta'} }{ \sum_{\lambda, \nu} \mathcal{M}_{{\bf p}\lambda \rightarrow {\bf q}\nu} \mathcal{M}^{*}_{{\bf p}\lambda \rightarrow {\bf q}\nu}}.
\end{equation}

We also consider a scenario where we start with a pure incoming state instead, which is of the form
\begin{equation}\label{eq:in_pure_state}
| \psi_{-\infty} \rangle = | {\bf p}, \lambda \rangle,
\end{equation}
with $|\lambda \rangle \in \{ LL,LR,RL,RR\}$ and evolves to
$| \psi_{\infty} \rangle = S |{\bf p}, \lambda \rangle$
after scattering. Then, the scattered and filtered state is
\begin{equation}\label{eq:out_state_pure}
| \psi_{\text{out}} \rangle = \frac{1}{\sum_{\nu} |\mathcal{M}_{{\bf p}\lambda \rightarrow {\bf q}\nu}|^2} \sum_{\eta}  \mathcal{M}_{{\bf p}\lambda \rightarrow {\bf q}\eta} |{\bf q}, \eta \rangle .
\end{equation}

In this paper, we will evaluate the matrix elements $\mathcal{M}_{{\bf p}\lambda \rightarrow {\bf q}\eta}$ at tree-level.

\subsection{\label{sec:Optimal_measurement_of_pure_states} Optimal measurement of pure states}
The set of eigenvectors $\{ |e_{\pm} \rangle \}$ of $L_{\sigma}$ (the SLD with respect to parameter $\sigma$) form an optimal measurement basis for which the QCRLB is saturated \cite{liu2016quantum}. For a pure quantum state $\rho = \rho^2=|\psi \rangle \langle \psi |$, the SLD has the simple form
\begin{align}
  L_{\sigma} &= 2\partial_{\sigma} \big( |\psi \rangle \langle \psi |\big) \nonumber \\
  & = 2\big(|\partial_{\sigma} \psi \rangle \langle \psi |+ |\psi \rangle \langle \partial_{\sigma} \psi |\big)\,.
\end{align}

Given the pure outgoing state $|\psi\rangle=|\psi_{\text{out}}\rangle$ (Eq. (\ref{eq:out_state_pure})) our aim is to find the optimal measurement basis for the estimation of the external parameters $p$ and $\theta$. For this purpose we let $\sigma \in \{p, \theta\}$ and decompose the state $|\partial_{\sigma}\psi \rangle$ in the orthonormal $\{|\psi \rangle, |\psi^{\perp} \rangle\}$ basis, which satisfies
\begin{equation}\label{fig:psi+psi_perp_condition}
  \langle \psi |\psi^{\perp} \rangle =0 \,.
\end{equation}
$|\partial_{\sigma} \psi \rangle$ is written as
\begin{equation}\label{fig:del_psi_decomposed}
  |\partial_{\sigma} \psi \rangle = \frac{ \alpha |\psi \rangle + \beta |\psi^{\perp} \rangle }{ \sqrt{ |\alpha|^2 + |\beta|^2 } } \,,
\end{equation}
where $\alpha = \langle \psi | \partial_{\sigma}\psi \rangle$, and $\beta = \langle \psi^{\perp} | \partial_{\sigma}\psi \rangle$. Hence, the decomposition of the SLD in the $\{|\psi \rangle, |\psi^{\perp} \rangle\}$ basis is
\begin{equation}
  L = 2\big( \alpha |\psi \rangle + \beta |\psi^{\perp} \rangle \big) \langle \psi | + 2|\psi \rangle \big( \alpha^* \langle \psi| + \beta^* \langle\psi^{\perp}| \big) \,,
\end{equation}
which can be written in matrix form as
\begin{equation}
  L = 2\begin{pmatrix}
      \alpha + \alpha^* & \beta^* \\[1pt]
      \beta & 0 \\[1pt]
      \end{pmatrix} 
  \,.
\end{equation}
The eigenvectors of this matrix are 
\begin{equation}\label{eq:e_1_optimal}
  |e_{\pm} \rangle = \frac{\alpha + \alpha^* \pm 2 \sqrt{|\alpha|^2 + |\beta|^2}}{2 \beta} |\psi \rangle + |\psi^{\perp} \rangle \,,
\end{equation}
where
\begin{align}
  |\psi^{\perp} \rangle &= \frac{\sqrt{ |\alpha|^2 + |\beta|^2 } \, |\partial_{\sigma} \psi \rangle - \alpha | \psi \rangle }{ \beta } \, .
\end{align}

We now consider an optimal projective measurement on $\rho$ described by $\Pi_{\pm} = |e_{\pm} \rangle \langle |e_{\pm} |$. The probability distribution associated with this measurement is given by the PDF
\begin{equation}
    p(e_{\pm}| \sigma) = |\langle e_{\pm} | \psi_{\text{out}} \rangle |^2,
\end{equation}
where $|e_{\pm} \rangle$ are the eigenvalues of $L_{\sigma}$. The CFI
\begin{equation}
    \text{I}^{(C)}_{\sigma,\textbf{e}} = \sum_{\pm} \frac{1}{p(e_{\pm}| \sigma)} \cdot \bigg( \frac{\partial p(e_{\pm}| \sigma)}{\partial \sigma} \bigg)^2
\end{equation}
associated with this PDF is equivalent to the QFI associated with $\rho$. This is compared with the case where, instead, a projective measurement $\Pi_{\lambda} = |\lambda \rangle \langle \lambda |$ is applied in the $\lambda \in \{LL, LR, RL, RR \}$ helicity/polarisation basis, inducing the PDF
\begin{equation}
    p(\lambda| \sigma) = |\langle \lambda | \psi_{\text{out}} \rangle |^2,
\end{equation}
which has an associated CFI given by
\begin{equation}
   \text{I}^{(C)}_{\sigma,\boldsymbol{\lambda}} = \sum_{\lambda} \frac{1}{p(\lambda| \sigma)} \cdot \bigg( \frac{\partial p(\lambda| \sigma)}{\partial \sigma} \bigg)^2.
\end{equation}

\subsection{Classical Fisher information of local measurements }\label{QE_in_QED_scattering}
Beside evaluating the absolute bounds determined by the quantum Fisher information matrix, we shall also examine the estimation performance of completely local schemes, based on projective measurements in the helicity/polarisation product basis $\{|LL\rangle,|LR\rangle,|RL\rangle,|RR\rangle\}$.
This will be done by considering the {\em classical} Fisher information with respect to the relevant parameters ($p$ and $\theta$) of the probability distribution obtained through the Born rule from the final scattered state and the local POVM above. We will limit ourselves to single-parameter Fisher informations in this case, which we shall denote with $I^{(C)}_{p,\boldsymbol{\lambda}}(p,\theta)$ and $I^{(C)}_{\theta,\boldsymbol{\lambda}}(p,\theta)$

Notice that the eigenvectors of Eq.~(\ref{eq:e_1_optimal}) in the pure state case and, more generally, the eigenvectors of the SLD operator which optimise the quantum estimation, are typically entangled, calling for non-local, entangled measurements as optimal schemes. Evaluating the quantities $I^{(C)}_{p,\boldsymbol{\lambda}}(p,\theta)$ and $I^{(C)}_{\theta,\boldsymbol{\lambda}}(p,\theta)$ is then a way to assess the efficacy of local -- more realistic and less demanding than global -- measurement schemes.

\section{\label{sec:Method} Method}
We considered electron-muon ($e^{-} \mu^{-} \rightarrow e^{-} \mu^{-}$), as well as electron-photon (Compton: $e^{-} \gamma \rightarrow e^{-} \gamma$) scattering, for both maximally mixed and pure initial states. We then modeled the scattering and four-momentum filtering of these states, as shown in Section \ref{QED_scattering}, treating them at tree level in the COM frame. The three external degrees of freedom considered initially were $p$, $\theta$ and $\phi$. For each scattering process, the $16$ scattering amplitudes $\mathcal{M}_{{\bf p}\lambda \rightarrow {\bf q}\eta}$ (for the $4$ different pre-scattering and post-scattering helicity configurations $\lambda, \eta \in \{ LL, LR, RL, RR\}$) were calculated at the lowest order of perturbation (i.e. at tree level). The calculated $\mathcal{M}_{p\lambda \rightarrow q\eta}$ values were then used to find the scattered and four-momentum filtered states, given by Eq. (\ref{filtered_state1}) in the mixed state case, and by Eq. (\ref{eq:out_state_pure}) in the pure state case. The QFIM terms with respect to $p$, $\theta$, and $\phi$ were then calculated using Eq. (\ref{eq:Fisher_multiparam_mixed}) in the mixed state case, and by Eq. (\ref{eq:Fisher_multiparam_single}) in the pure state case. For pure states in the single-parameter estimation case, the QFI terms were compared to the CFI terms resulting from a helicity/polarisation basis (non-optimal) measurement.
All calculations were done numerically over the following range of parameters: $p \in \left[ \Delta p, 5\right]\,\text{MeV}$, $\theta \in \left[ \Delta \theta, \pi\right]$, and $\phi \in \left[ 0, 2\pi\right)$ for electron-muon scattering and $p \in \left[ \Delta p, 5\right]\,\text{MeV}$, $\theta \in \left[ 0, \pi \right]$, and $\phi \in \left[ 0, 2\pi\right)$ for Compton scattering. Numerical steps of $\Delta p = 0.01 \, \text{MeV}$,  $\Delta \theta = \frac{\pi}{500}$, and $\Delta \phi = \frac{\pi}{500}$ were chosen. The $5\,\text{MeV}$ momentum upper bound was chosen as a sensible cut-off of the high-momentum scattering regime, where we observed less variation in the values of all computed quantities when compared to the low-momentum regime. The reason for restricting the $p$ and $\theta$ domains is the occurrence of an infrared (IR) divergence at $p = 0$ in Compton \cite{qiao2021overview} and electron-muon scattering calculations, as well as the t-channel divergence at $\theta = 0$ in electron-muon scattering.

As one should expect by symmetry, no dependence on $\phi$ remains in the output state, even for cases of definite initial helicity or polarisation. We have tested this numerically as a preliminary check on our evaluations and, henceforth, we can just focus on the estimation of $p$ and $\theta$, characterised by the symmetric $2\times 2$ matrix $\textbf{I}(p, \theta)$. Under such a rank reduction, Eq. (\ref{eq:qcrlb_general}) can be written down explicitly to obtain the following expression for the QCRLB:
\begin{equation} \label{eq:2_by_2_qcrlb}
\begin{pmatrix}
\text{Var} [\hat{p} ] & \text{Cov} [\hat{p},  \hat{\theta} ]\\[5pt]
\text{Cov} [ \hat{\theta} , \hat{p}] & \text{Var} [ \hat{\theta} ] 
\end{pmatrix} \geq \frac{1}{N \left( \textbf{I}_{pp}\textbf{I}_{\theta \theta} - \textbf{I}_{p \theta}^2 \right)}
\begin{pmatrix}
\textbf{I}_{\theta \theta} & -\textbf{I}_{p \theta}\\[5pt]
-\textbf{I}_{p\theta} & \textbf{I}_{p p} 
\end{pmatrix}, 
\end{equation}
Note that an experimental design which maximises the determinant in the denominator above -- and, in general, the determinant of the Fisher matrix --  is called D-optimal, as this results in a minimised covariance matrix. Using the diagonal terms of the minimised covariance matrix, we computed the signal-to-noise ratio (SNR) of the optimal estimator $\hat{\sigma}$ given an optimal-basis measurement of the scattered two-particle state we consider. We adapt the SNR definition from Ref. \cite{yuan2024signal} for $N=1$ measurement as
\begin{equation} \label{eq:snr_definition}
    \text{SNR}_{\sigma} = \frac{\hat{\sigma}}{\sqrt{\text{Var}[\hat{\sigma}]}}
\end{equation}
and we let $\sigma = p, \theta$. The single-parameter lower bound shown in Eq. (\ref{eq:var_x_fixed_y}), which can be extended to QET, was used to compute the optimal-estimator SNRs ($\text{SNR}_{p, \, \text{fixed} \, \theta}$ and $\text{SNR}_{\theta, \, \text{fixed} \, p}$) in single-parameter estimation. 
Note that, in this work, we have evaluated the SNR by using the true value of the parameter $\sigma$, rather than an actual estimator $\hat{\sigma}$ (bear in mind, however, that the latter does converge to the former for unbiased estimators).
Let us also recall that $\text{Var}[\hat{\sigma}]$ is just a value for a single measurement: dividing its square root by the square root of a sampling rate $f_s$, yields the error $\Delta_{\hat{\sigma}}$ as a function of the sampling time $t_s$. This can be written in terms of $\text{SNR}_{\sigma}$ as
\begin{equation}
\Delta_{\hat{\sigma}} = \frac{\hat{\sigma}}{\text{SNR}_{\sigma} \sqrt{f_s t_s}},
\end{equation}
which is just a rewriting of Eq.~(\ref{eq:2_by_2_qcrlb}) for optimal estimators and measurements of the parameter $\sigma$.

\section{\label{sec:Results} Results}
In this section, the numerical values of the following terms are presented as plots of $p$ and $\theta$: the three independent QFIM elements $\textbf{I}_{pp}(p, \theta)$, $\textbf{I}_{p\theta}(p, \theta)$, and $\textbf{I}_{\theta \theta}(p, \theta)$, as well as the covariance and SNR values of the optimal estimators $\hat{p}$ and $\hat{\theta}$ for an estimation in the optimal measurement basis $\{ |e_{\pm} \rangle \}$ shown in Eq. (\ref{eq:e_1_optimal}). In the case of single-parameter estimation based on pure-state measurements, we compare these results to the ones resulting from estimation in the local helicity/polarisation basis. 

We report our findings in full detail in the following sections, and then provide the reader with a summary in Sec.~\ref{summa}.
We work in natural units ($\hbar = c = 1$) throughout.

\subsection{\label{sec:Electron_muon_scattering} Electron-muon scattering}
\begin{figure}
     \includegraphics[width=4cm]{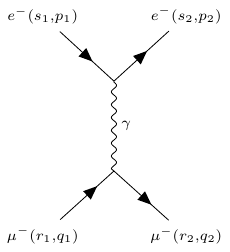}
     \caption{Feynman diagram of an electron-muon scattering
 process ($e^{-} \mu^{-} \rightarrow e^{-} \mu^{-}$).}
     \label{fig:e_mu_feynman}
\end{figure}

Here we present the QFIM elements as well as the covariance and SNR values of optimal estimators for the electron-muon scattering process ($e^{-} \mu^{-} \rightarrow e^{-} \mu^{-}$) with a Feynman diagram shown in Fig. \ref{fig:e_mu_feynman}. The scattering amplitude for this process is given by \cite{peskin_1995_introduction}
\begin{align}
i \mathcal{M} &= \bar{u}(s_2, p_2)(-ie \gamma^{\mu})u(s_1, p_1) \frac{-ig_{\mu \nu}}{(p_2 - p_1)^2}\nonumber\\
&\times \bar{u}(r_2, q_2)(-ie \gamma^{\nu})u(r_1, q_1),
\end{align}
where $u$ are Dirac spinors in the helicity basis, $\gamma^{\mu}$ are the Dirac matrices, and $g_{\mu \nu}$ is the Minkowski metric in flat spacetime for $\mu, \nu = 0,1,2,3$.

Results are shown for a scattered electron-muon pair state, given that the two particles were initially in a maximally mixed state (see Section \ref{sec:Electron_muon_scattering_mixed}) or in a pure state (see Section \ref{sec:Electron_muon_scattering_pure}).
%\FloatBarrier
\subsubsection{\label{sec:Electron_muon_scattering_mixed} Mixed state}
%\onecolumngrid\
\begin{figure*}
     \centering
     \includegraphics[width=\textwidth]
     %{compton_figs/mixed/all_mixed.pdf}
     {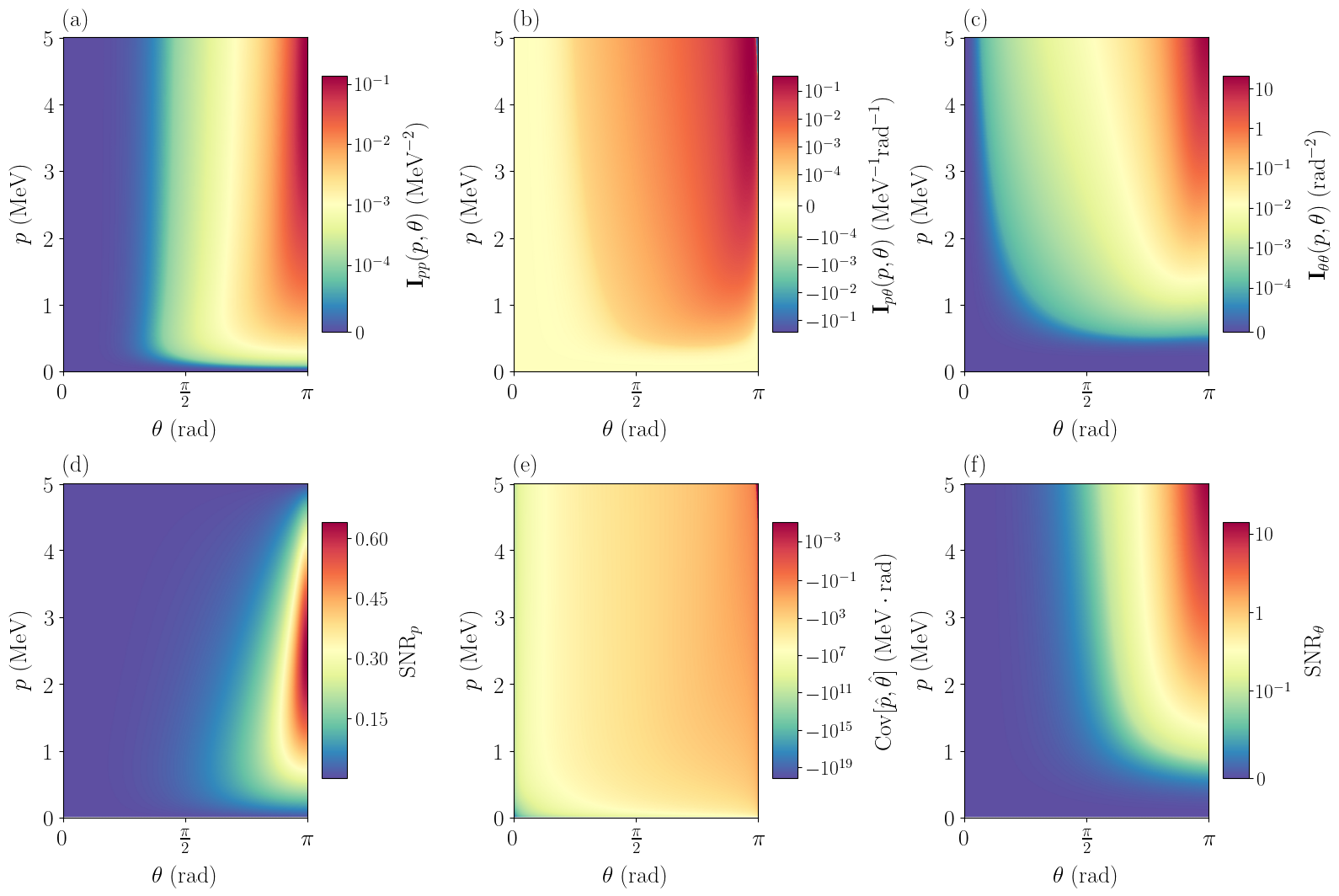}
     \caption{The Quantum Fisher information matrix elements $(a)$ $\textbf{I}_{pp}(p, \theta)$, $(b)$ $\textbf{I}_{p \theta}(p, \theta)$, and $(c)$ $\textbf{I}_{\theta \theta}(p, \theta)$ shown as functions of the particles' three-momentum magnitude $p$ and the polar scattering angle $\theta$ in electron-muon scattering. In the case of multi-parameter quantum estimation and for $N=1$ optimal-basis measurement, the signal-to-noise ratios $(\text{d})$ $\text{SNR}_p$ of the optimal estimator $\hat{p}$ and $(\text{f})$ $\text{SNR}_{\theta}$ of the optimal estimator $\hat{\theta}$ are shown, as well as the estimators' covariance $(\text{e})$ $\text{Cov} [ \hat{p}, \hat{\theta} ]$.}
    \label{fig:e_mu_I_mixed}
\end{figure*}
%\twocolumngrid\
A scattered maximally mixed electron-muon state has three unique QFIM elements $\textbf{I}_{pp}(p, \theta)$, $\textbf{I}_{p\theta}(p, \theta)$, and $\textbf{I}_{\theta \theta}(p, \theta)$ with similar features (see Fig. \ref{fig:e_mu_I_mixed} (a)-(c)). All three QFIM elements have consistently increasing values as $p$ and $\theta$ increase, i.e. the electron and muon helicities carry more information about $p$ and $\theta$ in a high-momentum, wide-angle scattering process. This means that, by selecting a back-scattered electron-muon pair, Alice can send more information about $p$ and $\theta$ to Bob. Heuristically, this complies with the loose intuition that the particles interact more strongly in this regime, thus making the final state very sensitive to initial parameters. In the low-$p$ scattering regime there is no $\theta$ dependence as the QFIM terms are negligible for all polar angles. For large $p$ the QFIM terms exhibit a much stronger $\theta$ dependence than a $p$ dependence, especially $\textbf{I}_{pp}(p, \theta)$, suggesting that the precisions of any unbiased $\hat{p}$ and $\hat{\theta}$ estimators must also have a weak $p$-dependence (especially $\hat{p}$). The largest QFI values are observed in the backscattering regime. For any given physical scenario (where the true $p$ and $\theta$ are fixed)  $\textbf{I}_{\theta \theta}(p, \theta)$ is larger than $\textbf{I}_{pp}(p, \theta)$, meaning that in general the electron and muon helicities carry more information about $\theta$ than $p$. For example, for large $p$ and $\theta$, $\textbf{I}_{\theta \theta}(p, \theta)$ is larger than $\textbf{I}_{pp}(p, \theta)$ by two orders of magnitude.

We now consider the covariance and SNR values of optimal $p$ and $\theta$ estimators for a single ($N=1$) optimal quantum measurement of the electron-muon state. The results for single-parameter and multi-parameter estimation are compared. The values of $\text{SNR}_{p}$ and $\text{SNR}_{\theta}$ (Fig. \ref{fig:e_mu_I_mixed} (d) and (f) respectively) both reach their maximum in the large $\theta$ range, as expected given the QFIM results. A significant difference in the features of the two quantities is that $\text{SNR}_{p}$ reaches a maximum around $p \approx 2.4 \text{MeV}$, whereas $\text{SNR}_{\theta}$ increases with $p$ in the considered domain. After a single optimal measurement of the mixed-state electron-muon helicities, the maximum $\text{SNR}_{p} \approx 0.6$ associated with the optimal estimator $\hat{p}$ suggests the standard deviation of $\hat{p}$ is consistently larger than the true value $p$. Hence, a reliable estimation of $p$ in such a scenario is not possible for $N=1$ and more measurements would be required. On the other hand, the maximum value of $\text{SNR}_{\theta} \approx 10$ for large $\theta$ suggests that a high-precision estimation of $\theta$ is possible even just for one optimal-basis measurement performed by Bob, provided that Alice has chosen a backscattered maximally mixed electron-muon pair. The precision of $\theta$ estimation in such a scenario is one order of magnitude larger than the precision of $p$ estimation.

As implied by Eq. (\ref{eq:var_x_fixed_y}), which also holds in QET, the QCRLBs of $\text{Var}[\hat{p}]_{\text{fixed}\, \theta}$ and $\text{Var}[\hat{\theta}]_{\text{fixed}\, p}$ (and hence the behaviour of $\text{SNR}_{p, \, \text{fixed} \, \theta}$ and $\text{SNR}_{\theta, \, \text{fixed} \, p}$) can be inferred from the behaviour of $\textbf{I}_{pp}(p, \theta)$ and $\textbf{I}_{\theta \theta}(p, \theta)$ respectively. Hence, these results are not shown explicitly here. No significant differences in the behaviour of $\text{SNR}_{p, \, \text{fixed} \, \theta}$ and $\text{SNR}_{\theta, \, \text{fixed} \, p}$ relative to $\text{SNR}_{p}$ and $\text{SNR}_{\theta}$ were observed.

Furthermore, as seen in Fig. \ref{fig:e_mu_I_mixed} (e), for $N=1$ measurement the saturated QCRLB of $\text{Cov}[\hat{p}, \hat{\theta}]$ is small and negative for all $p$ and for small $\theta$ values. $\text{Cov}[\hat{p}, \hat{\theta}]$ then gradually increases with $\theta$, until it becomes positive for $\theta \approx \pi$. This means that for an optimal-basis measurement, any pair of optimal unbiased $\hat{p}$ and $\hat{\theta}$ estimators must also exhibit a weak negative correlation for small $\theta$, which becomes a stronger negative correlation for larger $\theta$. The two are positively correlated for $\theta \approx \pi$. 

%\FloatBarrier
\subsubsection{\label{sec:Electron_muon_scattering_pure} Pure states}

Under a parity transformation $\hat{P}$, the two-particle helicity eigenstates transform as
$
|LL\rangle \overset{\hat{P}}{\leftrightarrow} |RR\rangle$ and 
$|LR\rangle \overset{\hat{P}}{\leftrightarrow} |RL\rangle$. The QED Lagrangian is invariant under parity transformation, so  the outgoing (scattered) pure quantum states transform as
$|\psi\rangle_{\text{out},LL} \overset{\hat{P}}{\leftrightarrow} |\psi\rangle_{\text{out},RR}$ and 
$|\psi\rangle_{\text{out},LR} \overset{\hat{P}}{\leftrightarrow} |\psi\rangle_{\text{out},RL}$,
up to relative phase terms, where $|\psi\rangle_{\text{out},\lambda}$ is the outgoing state for an incoming state $|\lambda\rangle$. The results shown in this section demonstrate that all pure-state QFIM elements are invariant under a parity transformation, as the QFIM for $|\psi\rangle_{\text{out},LL}$ equals the QFIM for $|\psi\rangle_{\text{out},RR}$ and the QFIM for $|\psi\rangle_{\text{out},LR}$ equals the one for $|\psi\rangle_{\text{out},RL}$. As a shorthand we will refer to the states $|LL\rangle$ and $|RR\rangle$ as ``same-helicity'' states, whereas $|LR\rangle$ and $|RL\rangle$ are referred to as ``opposite-helicity'' states. 

For pure-state electron-muon scattering, the three QFIM elements do not change significantly for opposite-helicity vs. same-helicity states. $\textbf{I}_{pp}(p, \theta)$ has higher values in the low-$p$ region, centred around $( p=0.52, \theta=1.91)$. Hence for all possible electron-muon pure states, the helicities of the two particles carry significant information about $\theta$ only in the low-$p$ scattering regime. The only difference between the $\textbf{I}_{pp}(p, \theta)$ functions for the two families of scattering electron-muon states is that for same-helicity states, $\textbf{I}_{pp}(p, \theta)$ has slightly larger values for large $p$ and $\theta$.

\begin{figure}
     \centering
     \includegraphics[width=9cm]
     %{e_mu_figs/pure/I_p_p.pdf}
     {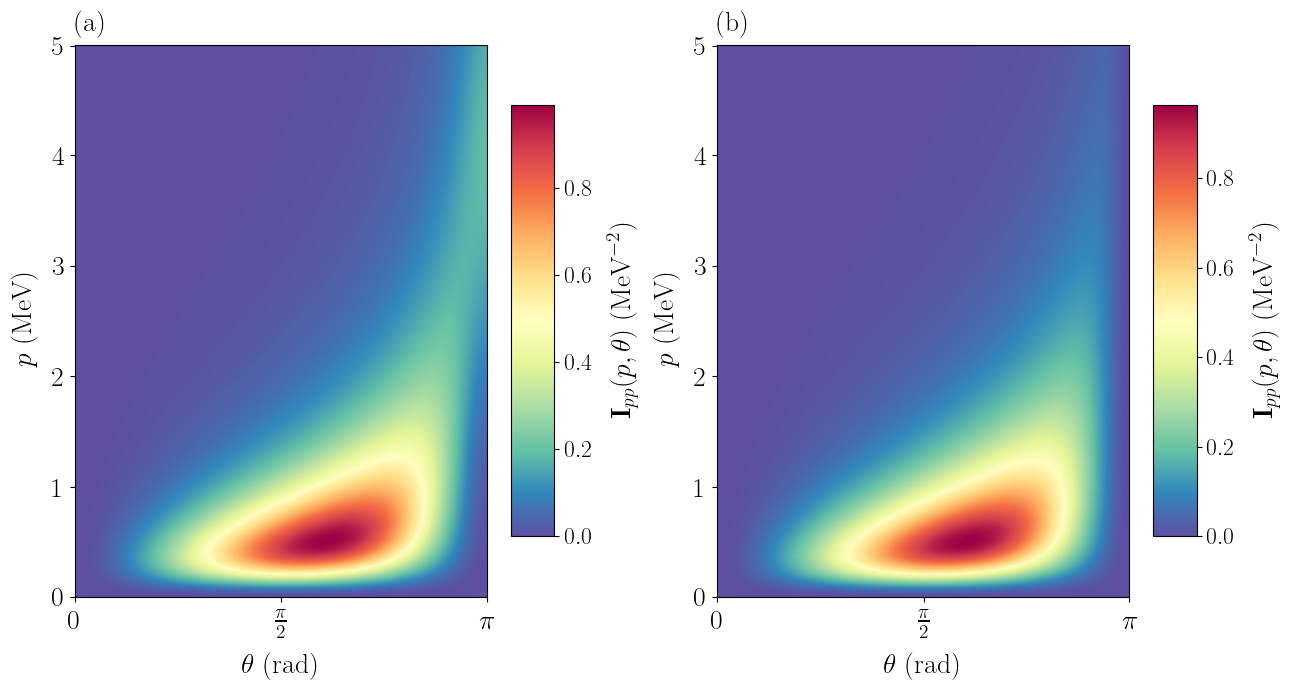}
     \caption{Quantum Fisher information matrix element $\textbf{I}_{pp}(p, \theta)$ with respect to the particles' three-momentum magnitude $p$ in electron-muon scattering. Results are shown for same-helicity pure incoming electron-muon states $|LL\rangle$ and $|RR\rangle$ in $(\text{a})$ and opposite-helicity states $|LR\rangle$ and $|RL\rangle$ in $(\text{b})$. These are plotted as functions of $p$ and the polar scattering angle $\theta$.}
     \label{fig:e_mu_I_p_p_pure}
\end{figure}

\begin{figure}
     \centering
     \includegraphics[width=9cm]
     %{e_mu_figs/pure/I_p_theta.pdf}
     {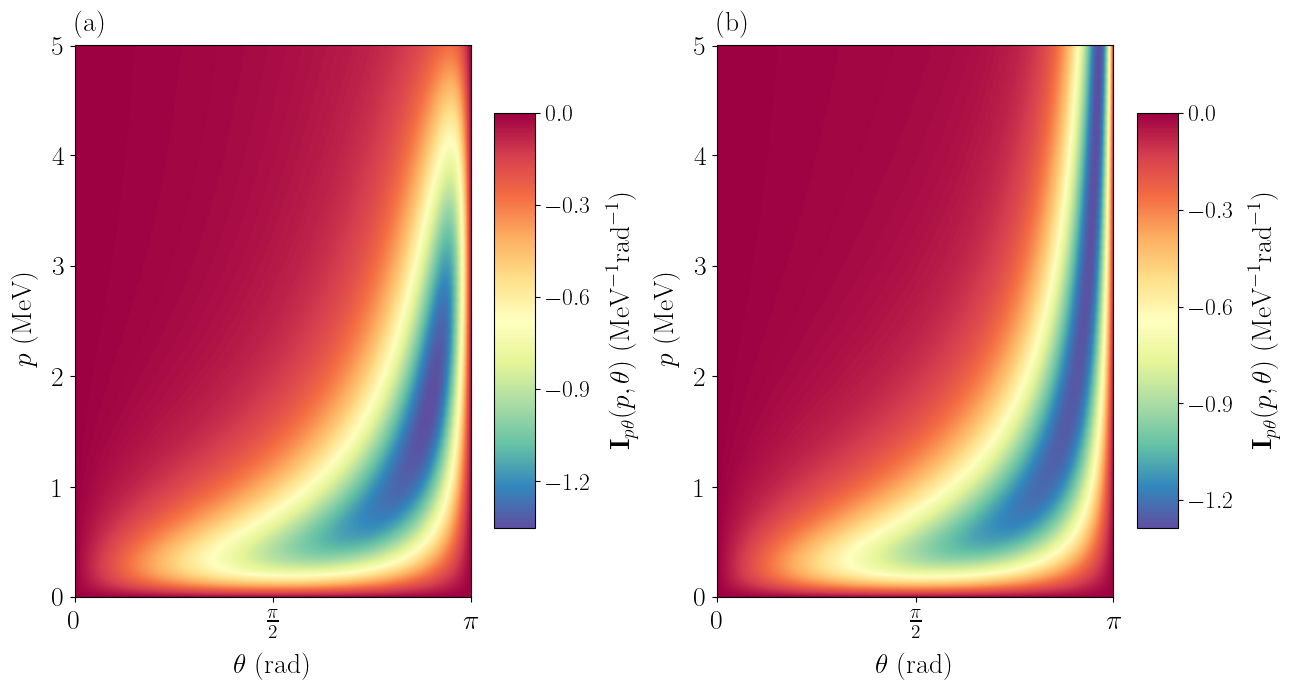}
     \caption{Quantum Fisher information matrix element $\textbf{I}_{p\theta}(p, \theta) = \textbf{I}_{\theta p}(p, \theta)$ with respect to the particles' three-momentum magnitude $p$ and the polar scattering angle $\theta$ in electron-muon scattering. Results are shown for same-helicity pure incoming electron-muon states $|LL\rangle$ and $|RR\rangle$ in $(\text{a})$ and opposite-helicity states $|LR\rangle$ and $|RL\rangle$ in $(\text{b})$. These are plotted as functions of $p$ and $\theta$.}
     \label{fig:e_mu_I_p_theta_pure}
\end{figure}

For both opposite-helicity and the same-helicity state scattering events $\textbf{I}_{p\theta}(p, \theta) \geq 0$ has a larger magnitude that is skewed in the $\frac{\pi}{2}<\theta<\pi$ region (see Fig. \ref{fig:e_mu_I_p_theta_pure}). The only difference between the two pure-state parity scenarios is that for the scattering of same-helicity pure states, a gradual decrease of the $\textbf{I}_{pp}(p, \theta)$ magnitude is seen again for $p>4\,\text{MeV}$, while no such decrease is seen for opposite-helicity pure states in the considered $p$ domain. As seen in Eq. (\ref{eq:2_by_2_qcrlb}), the off-diagonal QFIM terms affect the QCRLB on $\text{Cov} [ \hat{p}, \hat{\theta} ]$, which has a direct interpretation. As it will be shown, in the case of electron-muon scattering the $\textbf{I}_{p\theta}(p, \theta)$ values are consistently of a smaller magnitude than $\textbf{I}_{pp}(p, \theta)$ and $\textbf{I}_{\theta \theta}(p, \theta)$ and hence the former are dominated by the latter in the values of the $\text{Cov} [ \hat{p}, \hat{\theta} ]$ lower bound.

\begin{figure}
     \centering
     \includegraphics[width=9cm]
     %{e_mu_figs/pure/I_theta_theta.pdf}
     {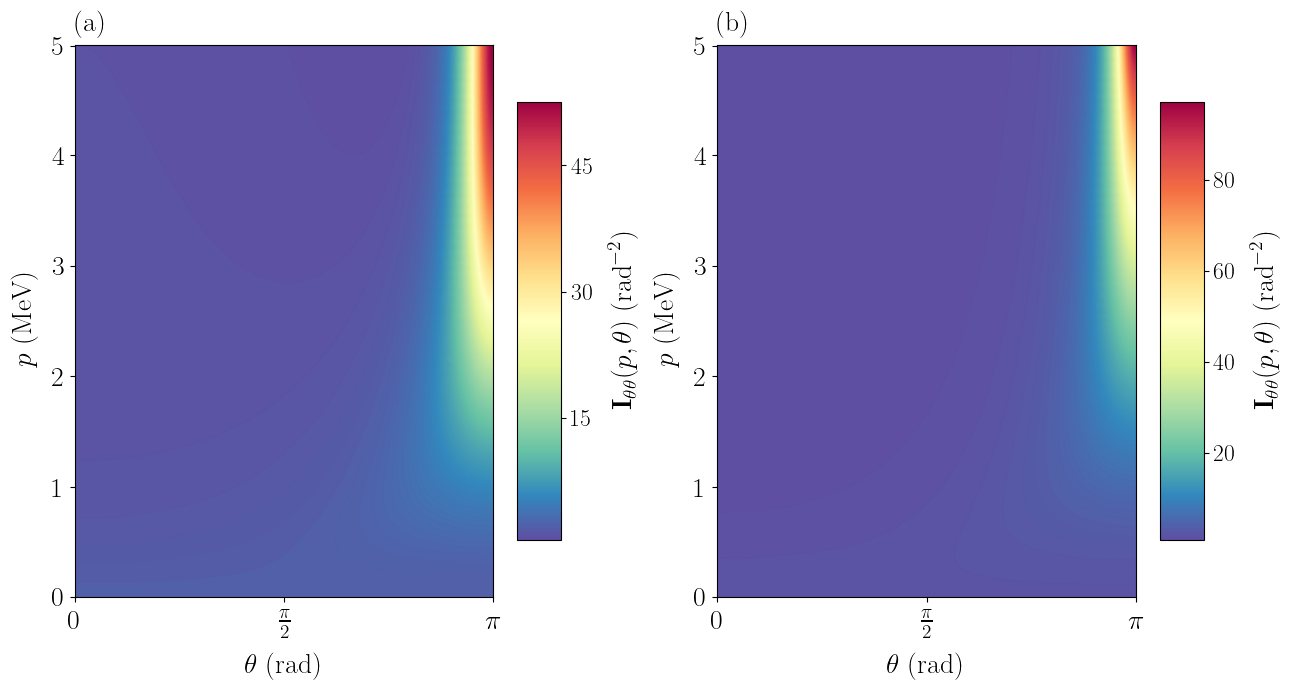}
     \caption{Quantum Fisher information matrix element $\textbf{I}_{\theta \theta}(p, \theta)$ with respect to the polar scattering angle $\theta$ in electron-muon scattering. Results are shown for same-helicity pure incoming electron-muon states $|LL\rangle$ and $|RR\rangle$ in $(\text{a})$ and opposite-helicity states $|LR\rangle$ and $|RL\rangle$ in $(\text{b})$. These are plotted as functions of the particles' centre-of-mass three-momentum $p$ and $\theta$.}
     \label{fig:e_mu_I_theta_theta_pure}
\end{figure}

As shown in Fig. \ref{fig:e_mu_I_theta_theta_pure} the $\textbf{I}_{\theta \theta}(p, \theta)$ QFIM term also has similar features for incoming same-helicity and opposite-helicity electron-muon states. Significant $\textbf{I}_{\theta \theta}(p, \theta)$ values are only seen in the region where $\theta$ approaches $\pi$, with larger values for larger $p$. Hence, after high-$p$ (and hence high-energy) backscattering, the electron and muon helicities carry more information about $\theta$. Even though $\textbf{I}_{\theta \theta}(p, \theta)$ reaches higher maxima than the two other unique QFIM elements, a significant $\textbf{I}_{\theta \theta}(p, \theta)$ value can only be achieved for a limited family of physical scenarios for pure-state electron-muon scattering. Furthermore, the maximum calculated information they can carry is larger if the two helicities were different before scattering. However, opposite-helicity incoming states reach a larger $\textbf{I}_{\theta \theta}(p, \theta)$ maximum for a smaller domain of $p$ and $\theta$ values distributed over larger $p$ and $\theta$ compared to the same-helicity state case. Comparing Fig. \ref{fig:e_mu_I_theta_theta_pure} with Fig. \ref{fig:e_mu_I_mixed} (c) shows that $\textbf{I}_{pp}(p, \theta)$ reaches a higher maxium for a pure, scattered electron-muon state than for a mixed state, albeit over a smaller $p$-$\theta$ domain.

We also computed the CFI terms $I^{(C)}_{p,\boldsymbol{\lambda}}(p,\theta)$ and $I^{(C)}_{\theta,\boldsymbol{\lambda}}(p,\theta)$ associated with a measurement of a pure, scattered electron-muon state in the electron-muon helicity basis. Invariance under state parity transformation was observed for the CFI terms as well. However, these terms are not shown here as they have the same features as those of $\textbf{I}_{pp}(p,\theta) = I^{(C)}_{p,\textbf{e}}(p,\theta)$ in Fig. \ref{fig:e_mu_I_p_p_pure} and $\textbf{I}_{\theta \theta}(p,\theta) = I^{(C)}_{\theta,\textbf{e}}(p,\theta)$ in Fig. \ref{fig:e_mu_I_theta_theta_pure}, respectively. Even though $I^{(C)}_{p,\boldsymbol{\lambda}}(p,\theta) < \textbf{I}_{pp}(p,\theta)$ as expected, the values of $I^{(C)}_{p,\boldsymbol{\lambda}}(p,\theta)$ and $\textbf{I}_{pp}(p,\theta)$ are of the same order of magnitude for both same-helicity and opposite-helicity electron-muon states. Therefore, when Bob is estimating $p$, he does not lose a significant amount of information by measuring Alice's electron-muon state in the helicity basis rather than the optimal measurement basis. This is in contrast with the $\theta$ estimation scenario, where the values of $I^{(C)}_{\theta,\boldsymbol{\lambda}}(p,\theta)$ are significantly smaller than those of $\textbf{I}_{\theta \theta}(p,\theta)$. More precisely, $\textbf{I}_{\theta \theta}(p,\theta) \in [ 0.51, 52.59 ] \,{\mathrm{rad}}^{-2}$ and $I^{(C)}_{\theta,\boldsymbol{\lambda}}(p,\theta) \in [0.50, 31.92 ] \,{\mathrm{rad}}^{-2}$ for same-helicity states, while $\textbf{I}_{\theta \theta}(p,\theta) \in [ 0.91, 97.32 ] \,{\mathrm{rad}}^{-2}$ and $I^{(C)}_{\theta,\boldsymbol{\lambda}}(p,\theta) \in [ 0.50, 56.90] \,{\mathrm{rad}}^{-2}$ for opposite-helicity states. Hence, a measurement of a prepared electron-muon state in the helicity basis of product states, rather than the optimal measurement basis, places a larger limit on the single-parameter estimation of $\theta$ than it does on the estimation of $p$. Additionally, we have evaluated the concurrence (a measure of entanglement easily evaluated for two-qubit states \cite{horodeckis}) of the optimal measurement-basis states $|e_{+}\rangle$ and $|e_{-}\rangle$ of Eq.~(\ref{eq:e_1_optimal}) and found that (in the low-$p$ scattering regime) each of the two states is less entangled in the helicity degrees of freedom in the case of $p$ estimation than in the case of $\theta$ estimation. As $I^{(C)}_{p,\boldsymbol{\lambda}}(p,\theta)$ and $\textbf{I}_{pp}(p,\theta)$ have significant values for small $p$, this finding complies with the fact that the CFI associated with standard, local helicity measurements is much closer to the actual QFI for $p$ than for $\theta$. 

\begin{figure}
     \centering
     \includegraphics[width=9cm]
     %{e_mu_figs/pure/Var_p.pdf}
     {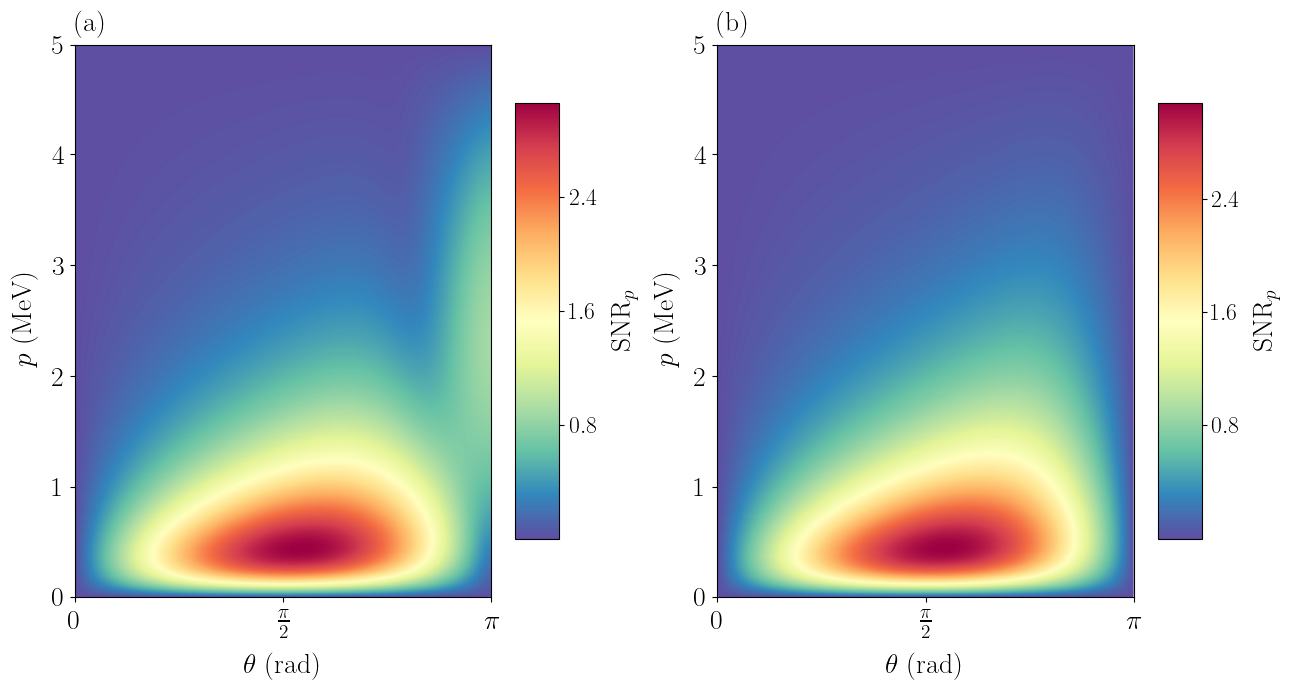}
     \caption{The signal-to-noise ratio $\text{SNR}_p$ of an optimal estimator $\hat{p}$ of the particles' three-momentum magnitude $p$ in electron-muon scattering. $N=1$ optimal-basis measurement of the electron-muon state is considered in the context of multi-parameter estimation.
     Results are shown for same-helicity pure incoming electron-muon states $|LL\rangle$ and $|RR\rangle$ in $(\text{a})$ and opposite-helicity states $|LR\rangle$ and $|RL\rangle$ in $(\text{b})$. Values are plotted as functions of $p$ and the polar scattering angle $\theta$.}
     \label{fig:e_mu_var_p_pure}
\end{figure}

We now compare the SNRs of optimal $\hat{p}$ and $\hat{\theta}$ estimators in three different scenarios: single vs. multi-parameter estimation for $N=1$ optimal measurement, and single-parameter estimation for one helicity/polarisation-basis measurement. As shown in Fig. \ref{fig:e_mu_var_p_pure}, $\text{SNR}_p$ has similar values for all electron-muon pure states, with a maximum of $\text{SNR}_p \approx 2.4$ for $\theta \approx \frac{\pi}{2}$ and $p \approx 0.5 \text{MeV}$. A distinguishing feature of same-helicity pure electron-muon states is a slightly larger $\text{SNR}_p$ for $\theta \approx \pi$ for a shown range of $p$ values. The maximum reached by $\text{SNR}_p$ for an optimal $\hat{p}$ and optimal-basis measurement of a pure state is four times larger than the maximum reached by $\text{SNR}_p$ in the mixed-state case, suggesting that in the former case a precise estimation of $p$ is possible in the low-$p$ region. However, in both cases $\text{SNR}_p < 1$ for a large domain of $p$ and $\theta$ values, making corresponding estimations (based on a single measurement) unreliable. In the single-parameter estimation scenario when $\theta$ is fixed to a known value, $\text{SNR}_{p, \, \text{fixed} \, \theta}$ has similar features to $\text{SNR}_{p}$, but reaches a higher maximum of $\text{SNR}_{p, \, \text{fixed} \, \theta} \approx 4$. Hence, if Bob knows the true value of $\theta$, the precision limit on his estimation of $p$ is improved. In the case of $p$ estimation, no significant differences are observed between the values of $\text{SNR}_{p, \, \text{fixed} \, \theta}$ resulting from an optimal-basis vs. a helicity basis measurement of the electron-muon state helicities.

\begin{figure}
     \centering
     \includegraphics[width=9cm]
     %{e_mu_figs/pure/Cov_p_theta.pdf}
     {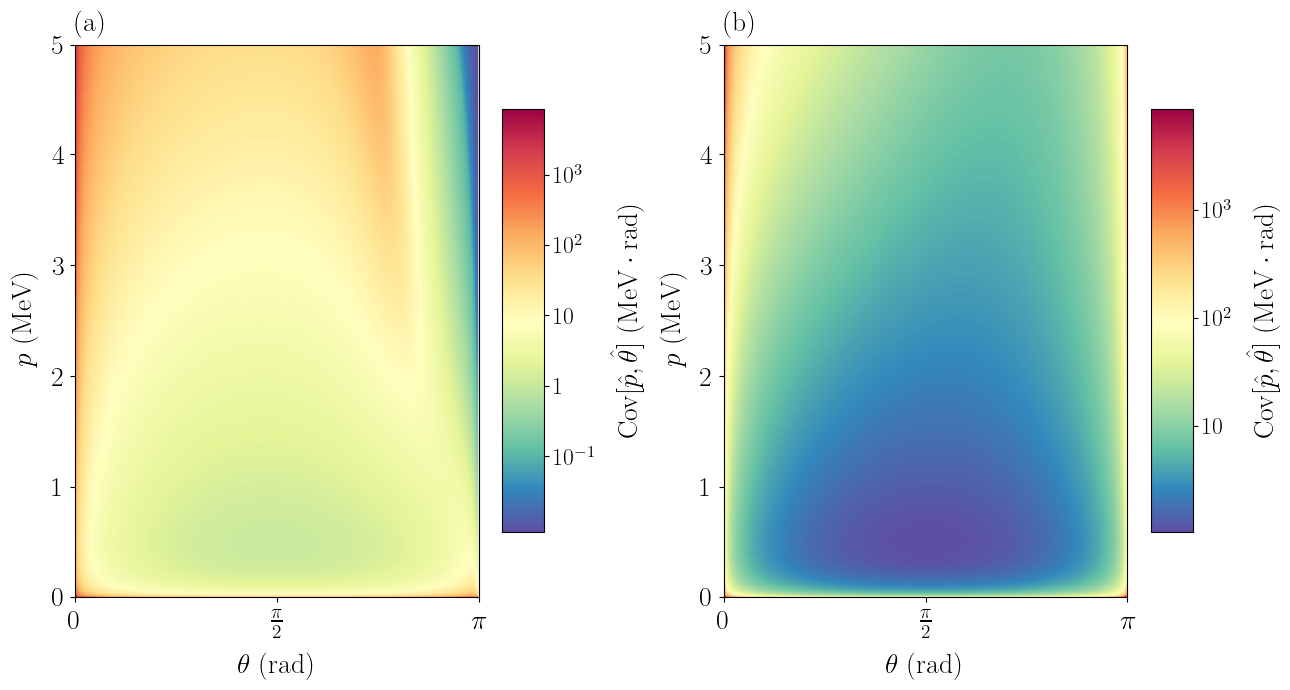}
     \caption{The lower bound of the covariance $\text{Cov} [ \hat{p}, \hat{\theta} ]$ of optimal estimators $\hat{p}$ and $\hat{\theta}$ of, respectively, the particles' three-momentum magnitude $p$ and the polar scattering angle $\theta$ in electron-muon scattering for $N=1$ optimal-basis measurement. Results are shown for same-helicity pure incoming electron-muon states $|LL\rangle$ and $|RR\rangle$ in $(\text{a})$ and opposite-helicity states $|LR\rangle$ and $|RL\rangle$ in $(\text{b})$. The variances are plotted as functions of $p$ and the polar scattering angle $\theta$.}
     \label{fig:e_mu_cov_p_theta_pure}
\end{figure}

$\text{Cov} [ \hat{p}, \hat{\theta} ]$, with respect to a pure state of a scattering electron-muon pair after $N=1$ optimal basis measurement, has a QCRLB that is positive for all $p$ and $\theta$, implying a positive correlation between the estimators $\hat{p}$ and $\hat{\theta}$ when $p$ and $\theta$ are both unknown (see Fig. \ref{fig:e_mu_cov_p_theta_pure}). This is in contrast with the distribution of $\text{Cov} [ \hat{p}, \hat{\theta} ]$ for a maximally mixed electron-muon state, which has a negative lower bound for most $p$ and $\theta$ values. Therefore, the correlation between $\hat{p}$ and $\hat{\theta}$ has been shown to depend on the purity of the electron-muon state. The maximum value of the QCRLB of $\text{Cov} [ \hat{p}, \hat{\theta} ]$ in the pure state case is $\approx 10^{3} \, \text{MeV} \cdot \text{rad}$ for both same-helicity and opposite-helicity states. Hence, the scale of $\text{Cov} [ \hat{p}, \hat{\theta} ]$ for pure-state electron-muon scattering is much larger in the pure-state case than in the mixed-state one, corresponding to a much stronger correlation between the estimators. The shape of the pure-state $p$-$\theta$ distribution of the optimal-measurement $\text{Cov} [ \hat{p}, \hat{\theta} ]$ QCRLB depends strongly on whether the helicities of the electron and muon are the same or not, even though the scale of the $\text{Cov} [ \hat{p}, \hat{\theta} ]$ QCRLB values does not.

\begin{figure}
     \centering
     \includegraphics[width=9cm]
     %{e_mu_figs/pure/Cov_p_theta.pdf}
     {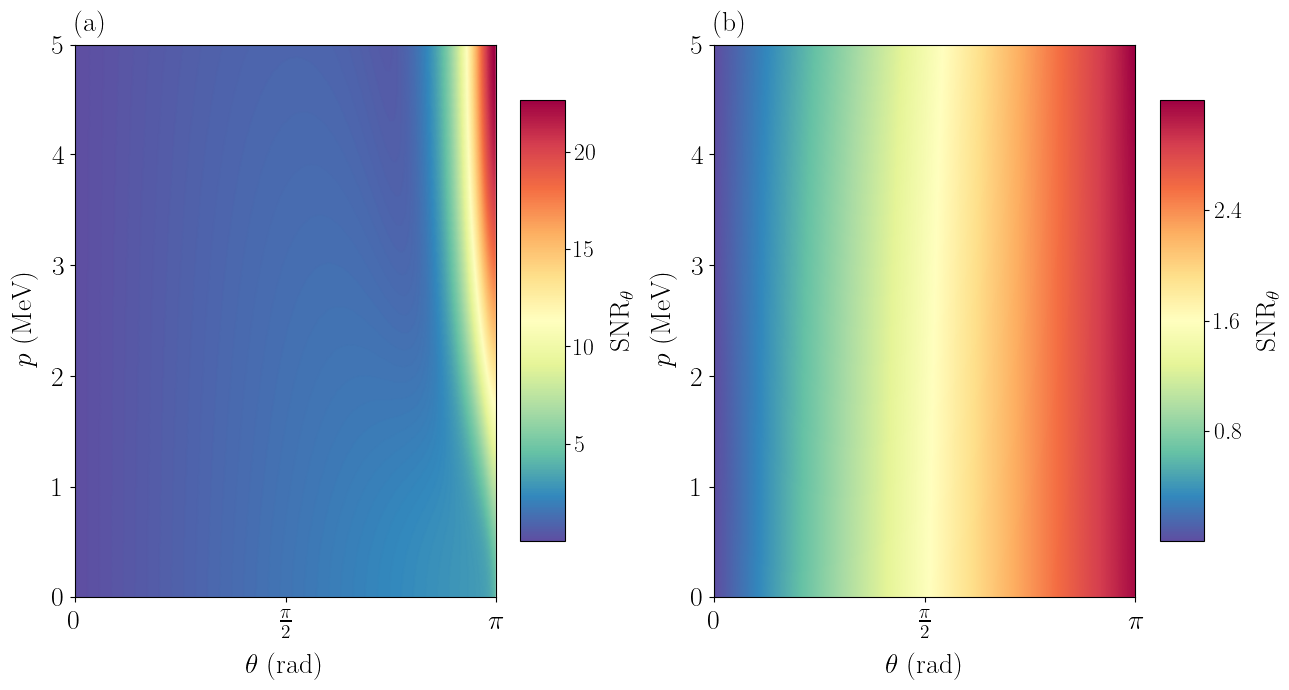}
     \caption{The signal-to-noise ratio $\text{SNR}_{\theta}$ of an optimal estimator $\hat{\theta}$ of the polar scattering angle in electron-muon scattering. $N=1$ optimal-basis measurement of the electron-muon state is considered in the context of multi-parameter estimation. Results are shown for same-helicity pure incoming electron-muon states $|LL\rangle$ and $|RR\rangle$ in $(\text{a})$ and opposite-helicity states $|LR\rangle$ and $|RL\rangle$ in $(\text{b})$. Values are plotted as functions of the particles' three-momentum magnitude $p$ and $\theta$.}
     \label{fig:e_mu_var_theta_pure}
\end{figure}

As shown in Fig. \ref{fig:e_mu_var_theta_pure}, in multi-parameter estimation $\text{SNR}_{\theta}$ due to one optimal-basis measurement of a pure electron-muon state shares the feature of mixed-state $\text{SNR}_{\theta}$ of reaching its maximum values in the large $\theta$ scattering regime. For same-helicity electron-muon states, $\text{SNR}_{\theta}$ reaches a maximum of $\approx 20$, corresponding to an estimation precision two times larger than in the equivalent mixed-state case, but applicable in a smaller domain of $(p,\theta)$ values. Note that this is the largest SNR reached in all different electron-muon scattering scenarios considered. On the other hand, in the opposite-helicity case, the maximum achievable $\text{SNR}_{\theta}$ is $\approx 2.4$, which is a decrease relative to the mixed-state case. However, in the former scenario, the domain of large $\text{SNR}_{\theta}$ extends over a larger domain of $\theta$ values. Now we consider the SNR when $\theta$ takes a known, fixed value, and the electron-muon state is still measured in the optimal basis. In the case of same-helicity states, no significant changes in the features and the value range of $\text{SNR}_{\theta, \, \text{fixed} \, p}$ are observed relative to $\text{SNR}_{\theta}$. However, for different electron-muon helicities, $\text{SNR}_{\theta, \, \text{fixed} \, p}$ reaches a maximum of $\approx 24$ (one order of magnitude higher than when $\theta$ is also estimated). Furthermore, the $(p,\theta)$ domain of large $\text{SNR}_{\theta, \, \text{fixed} \, p}$ values is restricted to the high-$\theta$ scattering regime, in a similar way to the same-helicity case. For single-parameter estimation based on a helicity/polarisation-basis measurement of the electron-muon state, the features of $\text{SNR}_{\theta, \, \text{fixed} \, p}$ remain identical to those in the optimal-basis measurement case, but the $\text{SNR}_{\theta, \, \text{fixed} \, p}$ values reach lower maxima. For large $\theta$ we computed $\text{SNR}_{\theta, \, \text{fixed} \, p} \approx 16$ for same-helicity states and $\text{SNR}_{\theta, \, \text{fixed} \, p} \approx 20$ for opposite-helicity states. Hence, in contrast to the $p$ estimation case, for $\theta$ estimation we observe a significant decrease in the precision of optimal estimation when a non-optimal measurement is performed.

%\FloatBarrier
\subsection{\label{subsec:Compton_scattering} Compton scattering}
\begin{figure}
     \centering
     \includegraphics[width=9cm]{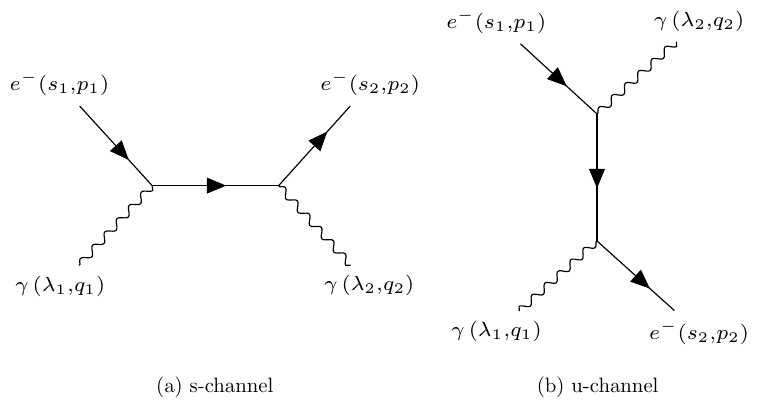}
     \caption{Feynman diagrams of the s and u channels for
 Compton scattering ($e^{-}\gamma \rightarrow e^{-}\gamma$).}
     \label{fig:compton_feynman}
\end{figure}

Here we present the QFIM elements and covariance matrix elements' lower bounds in Eq. (\ref{eq:2_by_2_qcrlb}) for Compton scattering ($e^{-} \gamma \rightarrow e^{-} \gamma$) with a Feynman diagram shown in Fig. \ref{fig:compton_feynman}. The scattering amplitude for this process is given by \cite{peskin_1995_introduction}
\begin{align}
   \mathcal{M} = \mathcal{M}_s + \mathcal{M}_u\,,
\end{align}
where the s-channel scattering amplitude is given by
\begin{align}
i \mathcal{M}_s &= \bar{u}(s_2, p_2) (-ie\gamma^{\mu}) \epsilon^{*}_{\mu} (\lambda_2, q_2) \frac{\slashed{p}_1 + \slashed{q}_1 + m_{e}}{(p_1+q_1)^2 - m^2_{e}} \nonumber\\
&\times (-ie\gamma^{\nu}) \epsilon_{\nu} (\lambda_1, q_1) u(s_1, p_1),
\end{align}
and the u-channel scattering amplitude is given by
\begin{align}
i \mathcal{M}_u &= \bar{u}(s_2, p_2) (-ie\gamma^{\nu}) \epsilon^{*}_{\nu} (\lambda_1, q_1) \frac{\slashed{p}_1 - \slashed{q}_2 + m_{e}}{(p_1 - q_2)^2 - m^2_{e}} \nonumber\\
&\times (-ie\gamma^{\mu}) \epsilon_{\mu} (\lambda_2, q_2) u(s_1, p_1),
\end{align}
where $m_e$ is the mass of an electron. The QFIM element functions are shown for a scattered electron-photon pair state, given that the two particles were initially in a maximally mixed state (see Section \ref{subsubsec:Compton_scattering_mixed}) or in a pure state (see Section \ref{subsubsec:Compton_scattering_pure}).

%\FloatBarrier
%\onecolumngrid 
\subsubsection{\label{subsubsec:Compton_scattering_mixed} Mixed states}
%\twocolumngrid 

%\onecolumngrid\
\begin{figure*}
     \centering
     \includegraphics[width=\textwidth]
     %{compton_figs/mixed/all_mixed.pdf}
     {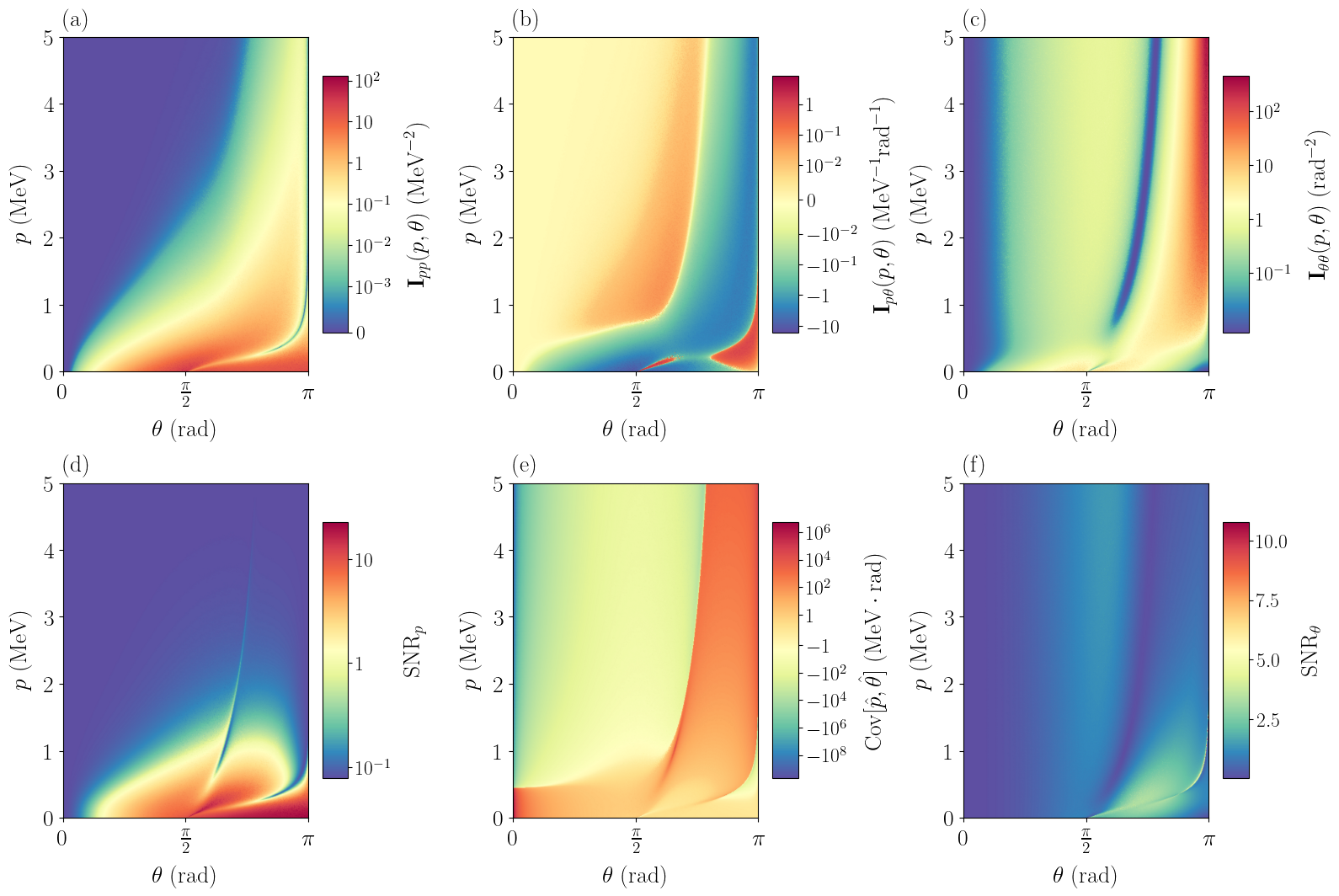}
     \caption{The Quantum Fisher information matrix elements $(a)$ $\textbf{I}_{pp}(p, \theta)$, $(b)$ $\textbf{I}_{p \theta}(p, \theta)$, and $(c)$ $\textbf{I}_{\theta \theta}(p, \theta)$ shown as functions of the particles' three-momentum magnitude $p$ and the polar scattering angle $\theta$ in Compton scattering. In the case of multi-parameter quantum estimation and for $N=1$ optimal-basis measurement, the signal-to-noise ratios $(\text{d})$ $\text{SNR}_p$ of the optimal estimator $\hat{p}$ and $(\text{f})$ $\text{SNR}_{\theta}$ of the optimal estimator $\hat{\theta}$ are shown, as well as the estimators' covariance $(\text{e})$ $\text{Cov} [ \hat{p}, \hat{\theta} ]$.}
    \label{fig:compton_I_mixed}
\end{figure*}
%\twocolumngrid\
In this section, we present Compton scattering results for an electron and a photon that are initially in a maximally mixed internal quantum state. The three unique QFIM elements for this scenario are shown in Fig. \ref{fig:compton_I_mixed} (a)-(c). The QFIM elements have functions of $p$ and $\theta$ with similar features. All three elements have small values in the region of small $\theta$. The size of this region increases for larger $p$. Therefore, for a maximally mixed electron-photon pair, the amount of information the internal degrees of freedom hold about the external degrees of freedom of the scattering is skewed with larger values for larger true $\theta$. The values of $\textbf{I}_{pp}(p, \theta)$ and $\textbf{I}_{\theta \theta}(p, \theta)$ both decrease locally at $\frac{\pi}{2}$ for small $p$. The line of small $\textbf{I}_{pp}(p, \theta)$ values also skews towards increasing $\theta$ as $p$ increases. $\textbf{I}_{p\theta}(p, \theta)$ shares a similar feature, with the difference that its values go from negative to positive around $\theta = \frac{\pi}{2}$ for small $p$. The maximum information of $p$ and $\theta$ carried by the scattered electron helicity and scattered photon polarisation is of the same order of magnitude as $10^2 \, \text{MeV}^{-2}$ for $p$ and  $10^2 \, \text{rad}^{-2}$ for $\theta$. This is larger than the information maxima that were observed in mixed-state electron-muon scattering. However, for Compton scattering $\textbf{I}_{pp}(p, \theta)$ has its largest values in the low-$p$ scattering regime, while $\textbf{I}_{\theta \theta}(p, \theta)$ is larger in the high-$\theta$, near-backscattering regime. The latter is a feature shared with electron-muon mixed-state scattering.

The covariance and SNR values of the optimal $\hat{p}$ and $\hat{\theta}$ (for one optimal-basis measurement) shown in Fig. \ref{fig:compton_I_mixed} (d)-(f) exhibit features that are similar to those of the relevant QFIM elements, with special behaviour at $\theta = \frac{\pi}{2}$ and in the $\theta \in [\frac{\pi}{2}, \pi]$ domain. $\text{SNR}_{p}$ and $\text{SNR}_{\theta}$ reach maxima that are of the same order of magnitude as $10$ (more precisely, $10.78$ for $\text{SNR}_{p}$ and $22.85$ for $\text{SNR}_{\theta}$) and both occur as $\theta \rightarrow \pi$. The difference between the two estimators is that the $\text{SNR}_{p}$ approaches its maximum value as $p \rightarrow 0$, while $\text{SNR}_{\theta}$ has its maximum value at $p=1.52$ for. A SNR value of the same order of magnitude as $10$ indicates that the optimal estimators in this scenario exhibit a high degree of precision. However, for most true $p$ and $\theta$ values, especially for high-$p$ scattering, $\text{SNR}<1$ and a reliable estimation based on one optimal measurement is not possible. Despite the need for multiple measurements, the mixed-state estimation precision for Compton scattering is overall higher than in the electron-muon scattering case, especially in the case of $p$ estimation. 

The values of these estimator $\text{SNRs}$ are significantly larger in the single-parameter estimation case. The general features of $\text{SNR}_{p, \, \text{fixed} \, \theta}$ and $\text{SNR}_{\theta, \, \text{fixed} \, p}$ are similar to those of $\text{SNR}_{p}$ and $\text{SNR}_{\theta}$, but the former have larger values for larger $p$ and $\theta$ domains than the latter. The maxima reached by $\text{SNR}_{p, \, \text{fixed} \, \theta}$ and $\text{SNR}_{\theta, \, \text{fixed} \, p}$ are $56.76$ and $66.54$ respectively, compared to $10.78$ for $\text{SNR}_{p}$ and $22.85$ for $\text{SNR}_{\theta}$. As expected, fixing one of the parameters improves the precision of the estimation of the other parameter.

As shown in Fig. \ref{fig:compton_I_mixed} (e), the maximum-magnitude values of the $\text{Cov} [ \hat{p}, \hat{\theta} ]$ QCRLB are $\text{Cov} [ \hat{p}, \hat{\theta} ] \sim 10^{6} \, \text{MeV}\cdot\text{rad}$ in the positive covariance region (for small $p$ and $\theta$ and for large $p$ and $\theta$) and $\text{Cov} [ \hat{p}, \hat{\theta} ] \sim -10^{8} \, \text{MeV}\cdot\text{rad}$ in the negative covariance region (for large $p$ and small $\theta$). In these true $p$-$\theta$ regions there is a strong correlation between the values estimated by $\hat{p}$ and $\hat{\theta}$ after an optimal basis measurement. However, for high $p$ in the $\theta \approx \frac{\pi}{2}$ region $\text{Cov} [ \hat{p}, \hat{\theta} ]$ is very small and the correlation between $\hat{p}$ and $\hat{\theta}$ is weak.
\FloatBarrier
\subsubsection{\label{subsubsec:Compton_scattering_pure} Pure states}
As one should expect, the invariance of QFI and CFI terms under parity transformation $\hat{P}$ is observed for pure-state Compton scattering, as it is for electron-muon scattering. Again, the CFI terms $I^{(C)}_{p,\boldsymbol{\lambda}}(p,\theta)$ and $I^{(C)}_{\theta,\boldsymbol{\lambda}}(p,\theta)$ (associated with a measurement of the pure electron-photon state in the helicity/polarisation basis) are not shown as they have the same shape as those of the QFIM terms $\textbf{I}_{pp}(p,\theta)$ and $\textbf{I}_{\theta \theta}(p,\theta)$, respectively. As expected, the CFI terms have consistently lower values. Notably, unlike in the electron-muon scattering case, the Fisher information terms for pure-state Compton scattering behave in significantly different ways for same-helicity and opposite-helicity states.

\begin{figure}[b]
     \centering
     \includegraphics[width=9cm]
     %{compton_figs/pure/I_p_p.pdf}
     {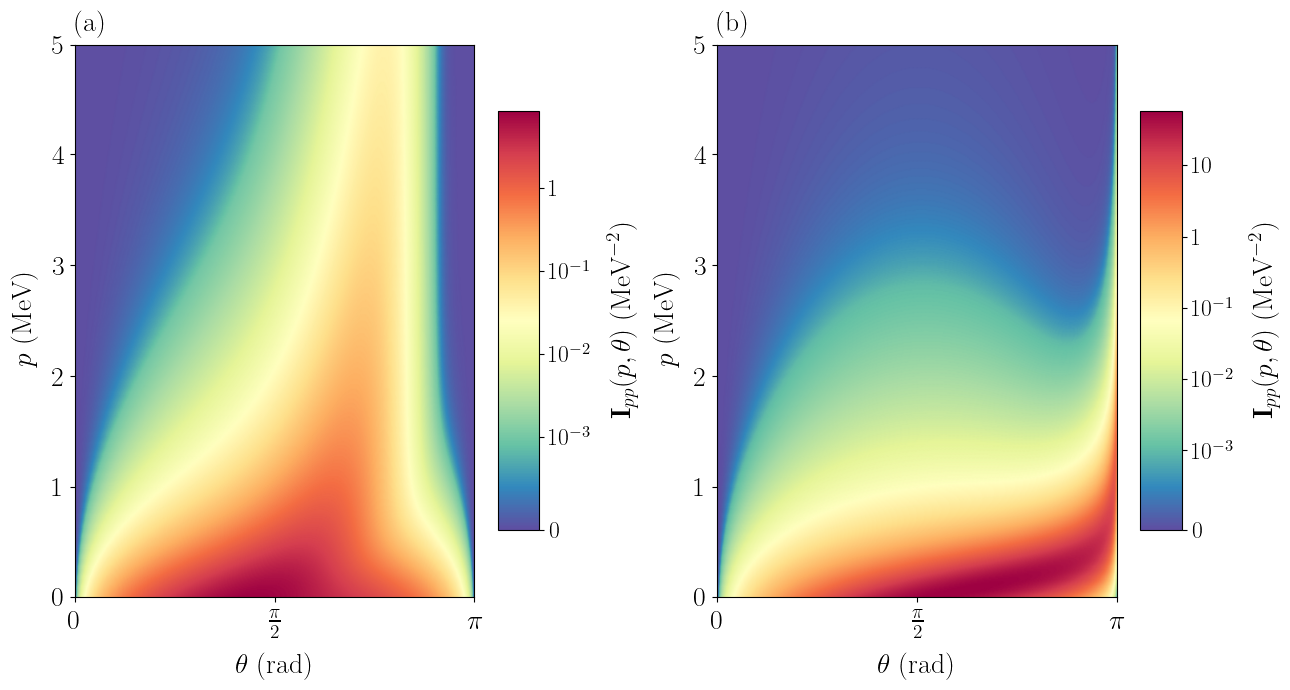}
     \caption{The signal-to-noise ratio $\text{SNR}_p$ of an optimal estimator $\hat{p}$ of the particles' three-momentum magnitude $p$ in Compton scattering. $N=1$ optimal-basis measurement of the electron-muon state is considered in the context of multi-parameter estimation.
     Results are shown for same-helicity pure incoming electron-photon states $|LL\rangle$ and $|RR\rangle$ in $(\text{a})$ and opposite-helicity states $|LR\rangle$ and $|RL\rangle$ in $(\text{b})$. Values are plotted as functions of $p$ and the polar scattering angle $\theta$.}
     \label{fig:compton_I_p_p_pure}
\end{figure}

As shown in Fig. \ref{fig:compton_I_p_p_pure}, the $\textbf{I}_{pp}(p, \theta)$ terms are both asymmetric about $\theta = \frac{\pi}{2}$ with higher values in the $\theta > \frac{\pi}{2}$ region. $\textbf{I}_{pp}(p, \theta)$ has a stronger $\theta > \frac{\pi}{2}$ skew for opposite-helicity pure states than for same-helicity ones. The largest calculated $\textbf{I}_{pp}(p, \theta)$ value is higher for the opposite-helicity scattered states than for same-helicity ones. In both parity cases however, $\textbf{I}_{pp}(p, \theta)$ values are higher for small $p$, and decrease for $\theta \approx 0$ and $\theta \approx \pi$. Therefore, the helicity and polarisation of the outgoing electron and photon can carry more information about a small $p$ value if the incoming helicity and polarisation  were different than if they were the same. By comparing Fig. \ref{fig:compton_I_p_p_pure} with Fig. \ref{fig:compton_I_mixed} (a) it is also noted that $\theta$, $\textbf{I}_{pp}(p, \theta)$  reaches a higher maximum value for a mixed scattered electron-muon state than for a pure state. This is the only QFIM term calculated in this paper for which this property is true. Furthermore, similarly to the electron-muon pure state scattering case, the values of $I^{(C)}_{p,\boldsymbol{\lambda}}(p,\theta)$ are of the same order of magnitude as those of $\textbf{I}_{pp}(p, \theta)$.

\begin{figure}[t!]
     \centering
     \includegraphics[width=9cm]
     %{compton_figs/pure/I_p_theta.pdf}
     {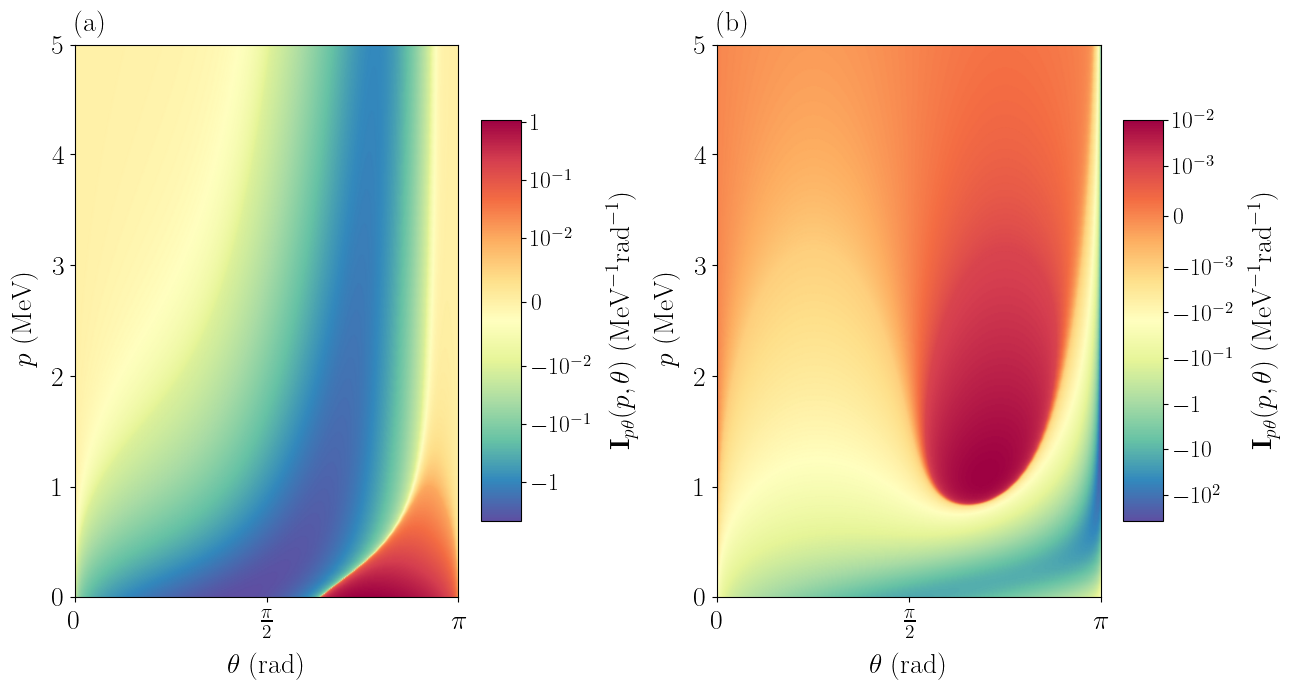}
     \caption{Quantum Fisher information matrix element $\textbf{I}_{p\theta}(p, \theta) = \textbf{I}_{\theta p}(p, \theta)$ with respect to the particles' three-momentum magnitude $p$ and the polar scattering angle $\theta$ in Compton scattering. Results are shown for same-helicity pure incoming electron-photon states $|LL\rangle$ and $|RR\rangle$ in $(\text{a})$ and opposite-helicity states $|LR\rangle$ and $|RL\rangle$ in $(\text{b})$ in the helicity-polarisation basis. These are plotted as functions of $p$ and $\theta$.}
     \label{fig:compton_I_p_theta_pure}
\end{figure}

The two pure-state $\textbf{I}_{p\theta}(p, \theta)$ functions are shown in Fig. \ref{fig:compton_I_p_theta_pure}. As already remarked, the physical interpretation of $\textbf{I}_{p\theta}(p, \theta)$ is understood by considering its effect on $\text{Cov} [ \hat{p}, \hat{\theta} ]$.
     
$\textbf{I}_{\theta \theta}(p, \theta)$ has significantly different values for same-helicity vs. opposite-helicity pure electron-photon states. For same-helicity states, large $\textbf{I}_{\theta \theta}(p, \theta)$ values only occur for a small domain of $p$ and $\theta$ values, where $\textbf{I}_{\theta \theta}(p, \theta)$ increases as $p$ and $\theta$ do, reaching a maximum of $\textbf{I}_{\theta \theta}(p, \theta) \sim 20 \,\text{rad}^{-2}$ in the considered domain. In contrast, for opposite-helicity states $\textbf{I}_{\theta \theta}(p, \theta) \sim 10^{4} \,\text{rad}^{-2}$ for $\theta \rightarrow \pi$. This is the largest order of magnitude value of a QFIM term calculated in this paper. The increase of $\textbf{I}_{\theta \theta}(p, \theta)$ for Compton backscattering events is a feature shared between opposite-helicity pure electron-photon states and mixed states, but does not hold for same-helicity states. Furthermore, for same-helicity states the values of $\textbf{I}_{\theta \theta}(p, \theta)$ and $I^{(C)}_{\theta,\boldsymbol{\lambda}}(p,\theta)$ are of the same order of magnitude. However, for opposite-helicity states $I^{(C)}_{\theta,\boldsymbol{\lambda}}(p,\theta)$ reaches a maximum value that is two orders of magnitude smaller than the one reached by $\textbf{I}_{\theta \theta}(p, \theta)$. More precisely, $I^{(C)}_{\theta,\boldsymbol{\lambda}}(p,\theta) \in [0.00, 569.78] \, \text{rad}^{-2}$ and $\textbf{I}_{\theta \theta}(p, \theta) \in [0.00, 12021.89] \, \text{rad}^{-2}$ for opposite-helicity states. This result is different from the electron-muon scattering result, where lower Fisher information was observed when measuring both same-helicity and opposite-helicity states in the helicity/polarisation basis.

\begin{figure}[t!]
     \centering
    \includegraphics[width=9cm]
    %{compton_figs/pure/I_theta_theta.pdf}
    {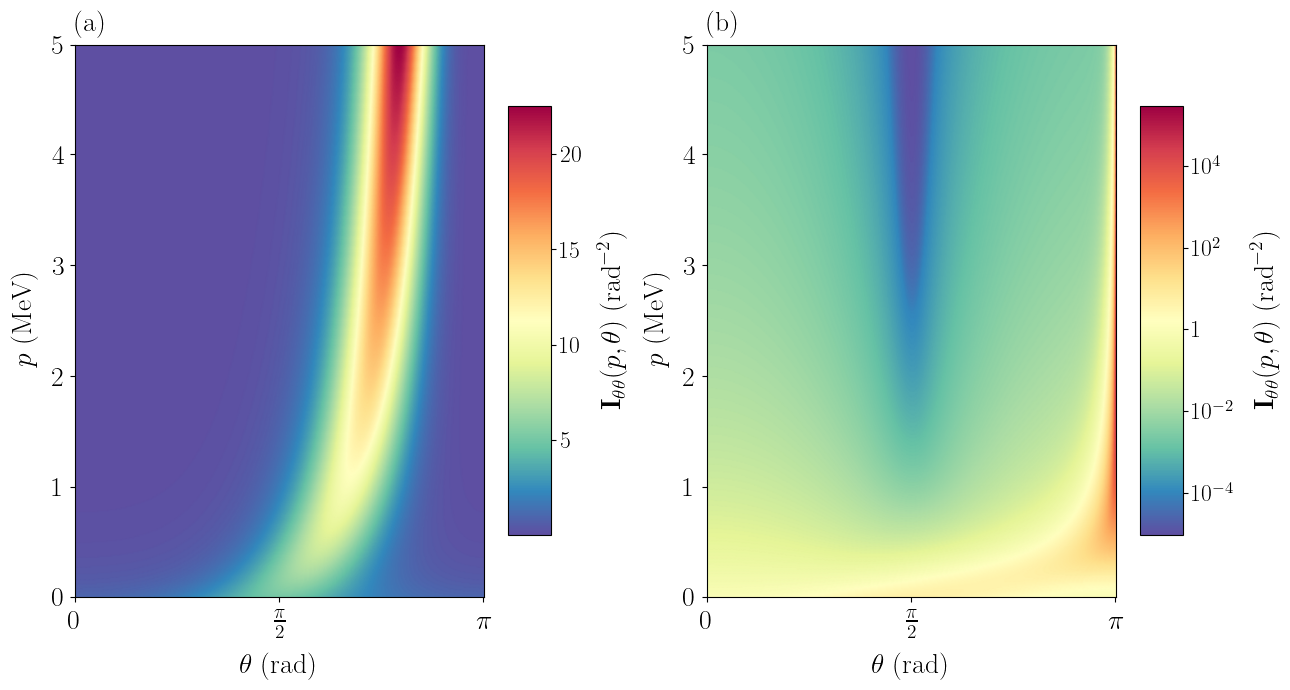}
     \caption{Quantum Fisher information matrix element $\textbf{I}_{\theta \theta}(p, \theta)$ with respect to the polar scattering angle $\theta$ in Compton scattering. Results are shown for same-helicity pure incoming electron-photon states $|LL\rangle$ and $|RR\rangle$ in $(\text{a})$ and opposite-helicity states $|LR\rangle$ and $|RL\rangle$ in $(\text{b})$ in the helicity-polarisation basis. These are plotted as functions of the particles' centre-of-mass three-momentum $p$ and $\theta$.}
     \label{fig:compton_I_theta_theta_pure}
\end{figure}

\begin{figure}
     \centering
     \includegraphics[width=9cm]
     %{compton_figs/pure/Var_p.pdf}
     {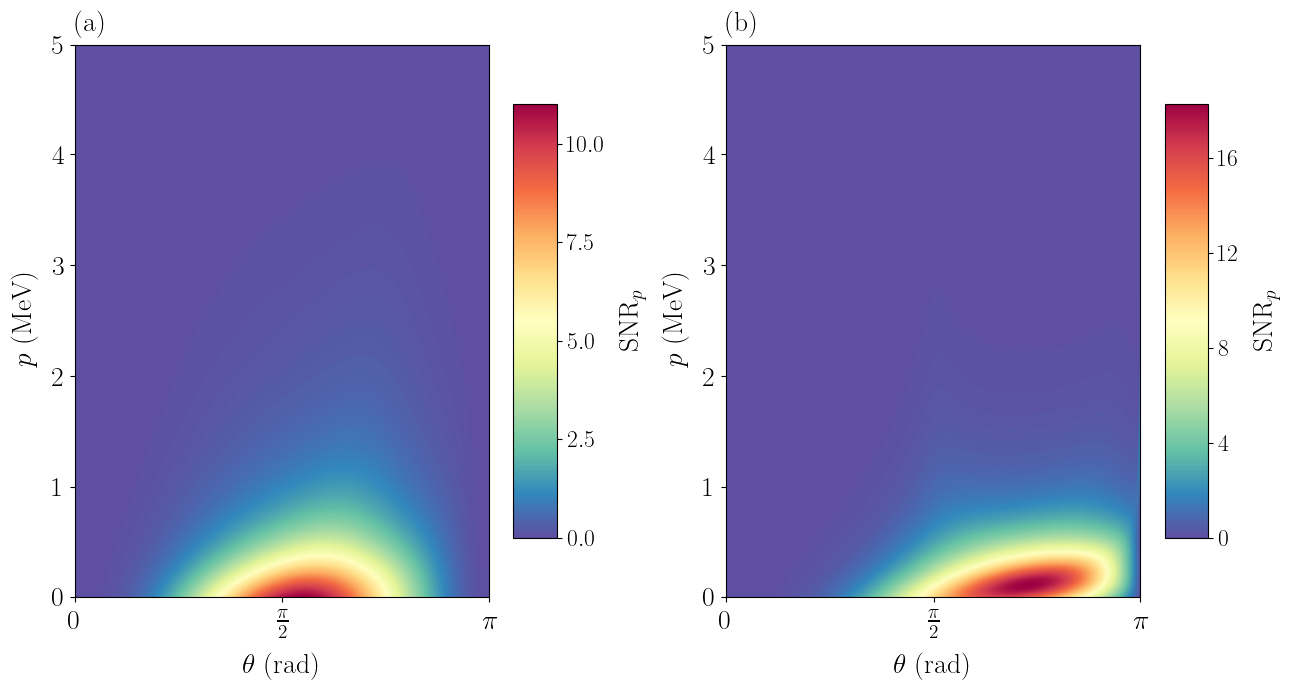}
     \caption{The lower bound of the variance $\text{Var} [ \hat{p} ]$ of an optimal estimator $\hat{p}$ of the particles' three-momentum magnitude $p$ in Compton scattering for an optimal-basis measurement. Results are shown for same-helicity pure incoming electron-photon states $|LL\rangle$ and $|RR\rangle$ in $(\text{a})$ and opposite-helicity states $|LR\rangle$ and $|RL\rangle$ in $(\text{b})$, where an estimation based on $N=1$ number of measurements was assumed. The variances are plotted as functions of $p$ and the polar scattering angle $\theta$.}
     \label{fig:compton_var_p_pure}
\end{figure}

We now analyse the behaviour of optimal $\hat{p}$ and $\hat{\theta}$ estimators in single and multi-parameter estimation for one optimal measurement, as well as in single-parameter estimation for one helicity/polarisation-basis measurement. As shown in Fig. \ref{fig:compton_var_p_pure}, given an optimal-basis measurement of a pure electron-photon state, a high-precision estimate of $p$ is possible only for low-$p$ Compton scattering. Large $\text{SNR}_p$ values skew towards the $\theta \in [\frac{\pi}{2}, \pi]$ domain, more so for opposite-helicity electron-photon states. The maximum reached by $\text{SNR}_p$ is $18.29$ for opposite-helicity pure states, which is larger than the maximum of $11.02$ for same-helicity states. In the single-parameter estimation regime when $\theta$ is know, $\text{SNR}_{p, \, \text{fixed} \, \theta}$ reaches higher values for a larger domain of $\theta$ values, while other features, such as the low precision for high $p$ scattering, remain the same as in the multi-parameter estimation case. The increase in $\text{SNR}_{p, \, \text{fixed} \, \theta}$ relative to $\text{SNR}_p$ is more significant for opposite-helicity electron-photon states, where $\text{SNR}_{p, \, \text{fixed} \, \theta}$ reaches a maximum of $37.41$. In the same-helicity case the maximum $\text{SNR}_{p, \, \text{fixed} \, \theta}$ value is $14.34$. When the pure electron-photon state is measured in the helicity/polarisation basis instead, the resulting $\text{SNR}_{p, \, \text{fixed} \, \theta}$ shares the features of $\text{SNR}_{p, \, \text{fixed} \, \theta}$ associated with an optimal-basis measurement. The values of the two SNRs are of the same order of magnitude, as expected due to the similarity between $I^{(C)}_{p,\boldsymbol{\lambda}}(p,\theta)$ and $\textbf{I}_{pp}(p, \theta)$. Hence, performing a single (non-optimal) helicity/polarisation-basis measurement leads to no significant decrease in the achievable $p$ estimation precision relative to the optimal measurement case. This is the same behaviour as the one observed in electron-muon scattering.

\begin{figure}
     \centering
     \includegraphics[width=9cm]
     %{compton_figs/pure/Cov_p_theta.pdf}
     {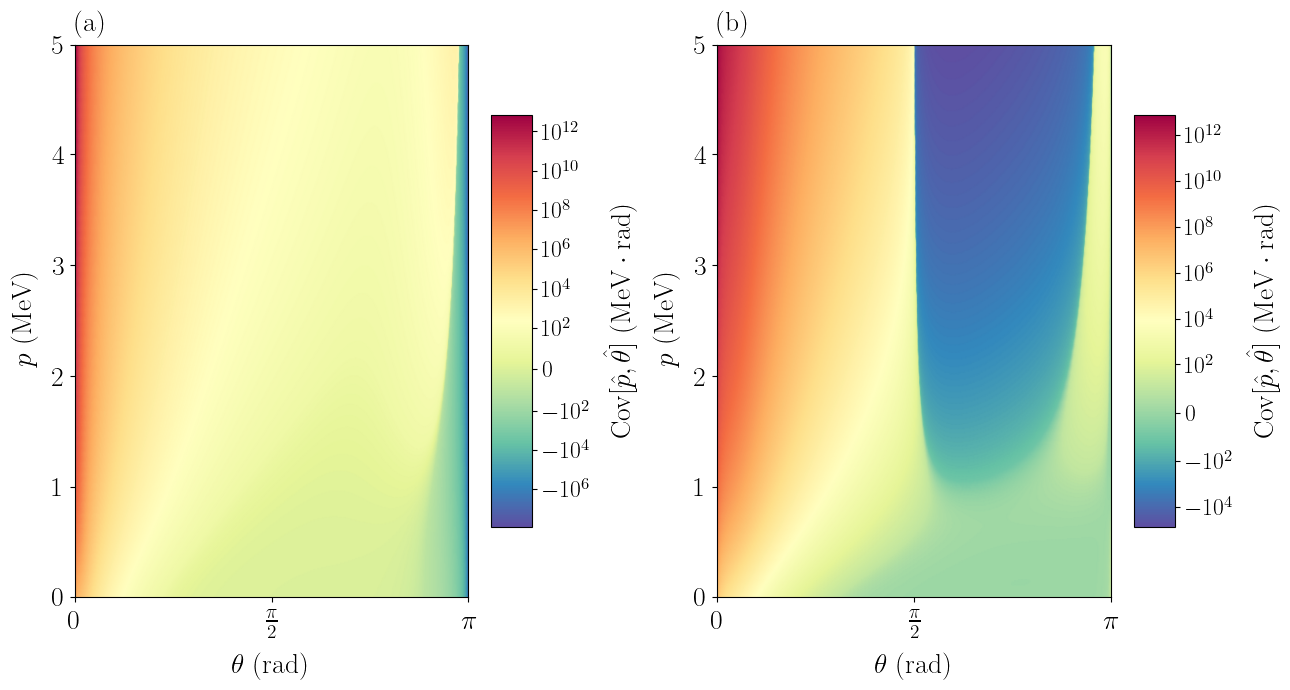}
     \caption{The lower bound of the covariance $\text{Cov} [ \hat{p}, \hat{\theta} ]$ of the optimal estimators $\hat{p}$ and $\hat{\theta}$ of, respectively, the particles' three-momentum magnitude $p$ and polar scattering angle $\theta$ in Compton scattering for $N=1$ optimal-basis measurement. Results are shown for same-helicity pure incoming electron-photon states $|LL\rangle$ and $|RR\rangle$ in $(\text{a})$ and opposite-helicity states $|LR\rangle$ and $|RL\rangle$ in $(\text{b})$. The variances are plotted as functions of $p$ and the polar scattering angle $\theta$.}
     \label{fig:compton_cov_p_theta_pure}
\end{figure}

As shown in Fig. \ref{fig:compton_cov_p_theta_pure}, the largest positive $\text{Cov} [ \hat{p}, \hat{\theta} ]$ QCRLB values in the considered $p$-$\theta$ domain are observed for small $\theta$ and increase with $p$. This corresponds to a strong correlation between the $\hat{p}$ and $\hat{\theta}$ estimators in the small $\theta$ scattering regime. The $\text{Cov} [ \hat{p}, \hat{\theta} ]$ QCRLB functions also have values close to $0$ in the small $p$ domain for $\theta$ near $\frac{\pi}{2}$. A large negative covariance is observed for $\theta \rightarrow \pi$ in the same-helicity state case, and for the domain $\theta \in [\frac{\pi}{2}, \pi]$ in the opposite-helicity case. The strength of this negative estimator correlation increases with $p$.

\begin{figure}[t!]
     \centering
    \includegraphics[width=9cm]
    %{compton_figs/pure/Var_theta.pdf}
    {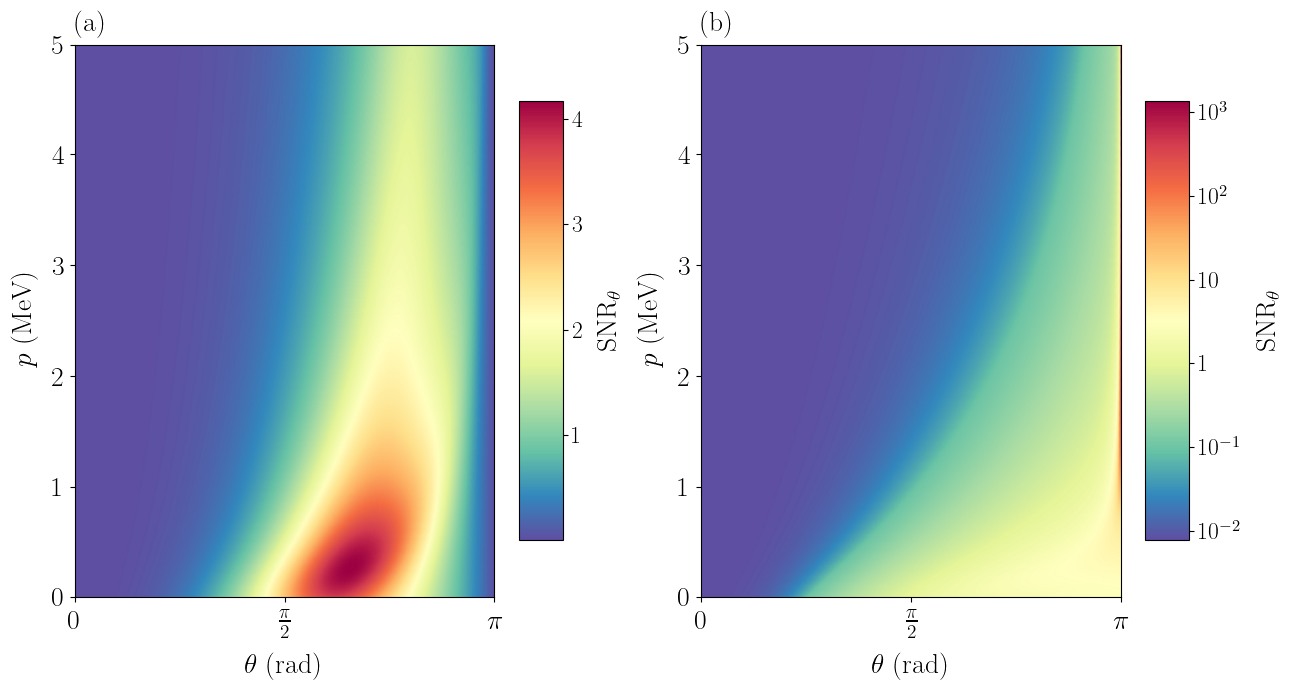}
     \caption{The signal-to-noise ratio $\text{SNR}_{\theta}$ of an optimal estimator $\hat{\theta}$ of the polar scattering angle in Compton scattering. $N=1$ optimal-basis measurement of the electron-photon state is considered in the context of multi-parameter estimation. Results are shown for same-helicity pure incoming electron-photon states $|LL\rangle$ and $|RR\rangle$ in $(\text{a})$ and opposite-helicity states $|LR\rangle$ and $|RL\rangle$ in $(\text{b})$. Values are plotted as functions of the particles' three-momentum magnitude $p$ and $\theta$.}
     \label{fig:compton_var_theta_pure}
\end{figure}

In Fig. \ref{fig:compton_var_theta_pure} we show the $\text{SNR}_{\theta}$ of an optimal $\hat{\theta}$ estimator for a single optimal-basis measurement of a pure, Compton-scattered electron-photon state. A significant difference is observed between the $\text{SNR}_{\theta}$ values for same-helicity and opposite-helicity electron-photon states. For same-helicity, significant $\theta$ estimation precision is achievable for the low-$p$, $\theta \in [\frac{\pi}{2}, \pi]$ domain of high $\text{SNR}_{\theta}$ values. $\text{SNR}_{\theta}$ reaches a maximum of $\sim 4$. For opposite-helicity states however, $\text{SNR}_{\theta} \sim 10^3$ is observed for high-$p$ Compton backscattering, i.e. in the limit $\theta \rightarrow \pi$. This is the highest optimal estimation precision (for just one measurement) observed in this paper. The relevant physical scenario is one where Alice prepares an electron and a photon with opposite helicity and circular polarisation values respectively, then lets them scatter and selects a backscattered two-particle state to send to Bob. Measuring the received state in the optimal basis, Bob can map his measurement result to an estimate of the scattering angle $\theta$. Just for one such measurement, he would be able to estimate $\theta$ to a very high precision. %However, as noted, due to the IR divergence of Compton scattering matrix elements at $\theta = \pi$, the value of $\text{SNR}_{\theta}$ was computed for $\theta \in [0, \pi - d\theta]$. Our numerical results indicate that the SNR might actually diverge around the infra-red divergence, which is an interesting feature per se.

We now consider single-parameter $\theta$ estimation in the scenario where $p$ is fixed but an optimal-basis measurement of the scattered electron-photon state is still performed. $\text{SNR}_{\theta, \, \text{fixed} \, p}$ share common features with $\text{SNR}_{\theta}$ for both same-helicity and opposite-helicity states. It is only for same-helicity states that large $\text{SNR}_{\theta, \, \text{fixed} \, p}$ occur for large $p$ (rather than for small $p$ as is the case for $\text{SNR}_{\theta}$). The maximum value of $\text{SNR}_{\theta, \, \text{fixed} \, p}$ in the same-helicity case is $\approx 10$. When a measurement is taken in the helicity/polarisation basis instead, a decrease in the values of $\text{SNR}_{\theta, \, \text{fixed} \, p}$ is observed, even though the features of the functions remain unchanged relative to the optimal-basis measurement, single-parameter estimation case. For a  helicity/polarisation basis measurement, the maximum of $\text{SNR}_{\theta, \, \text{fixed} \, p}$ (reached as $\theta \rightarrow \pi$) is $\sim 10^2$. As for electron-muon scattering, we observe a significant dependence of the precision of $\theta$ estimation on the measurement basis chosen.

\subsection{Summary of results}\label{summa}

Let us summarise here the salient features of our study:

\begin{itemize}

\item For electron-muon scattering and mixed initial states, optimal sensitivity to and estimation of the external parameters is obtained for backward scattering at high energies: if Alice intended to send Bob information about external parameters through the internal degrees of freedom, she would do well to select backscattered particles.

\item The behaviour above is common to the estimation of the scattering angle with initial pure product states, whilst the estimation of $p$ with pure initial states display a more sophisticate structure, with optimal angle between $\pi/2$ and $\pi$ and better resolution at lower energies (around the electron mass).

\item Final measurements in the local helicity basis are close to optimal to estimate the centre-of-mass momentum but they are largely suboptimal in determining the scattering angle $\theta$, which would profit from measurements in an entangled basis.

\item Compton scattering estimation has a more structured response to initial energies, achieving optimal estimation below 1 MeV and for a wider range of angles, although wider angles are still optimal. 

\item Also, in Compton scattering the performance of same-helicity and opposite-helicity initial pure states is very distinct, with the latter preferring lower energies and wider angles, and the former maintaining higher sensitivity for a wider range of energies.

\item For Compton scattering, the highest achievable precision we computed is the one for $\theta$ estimation based on an optimal-basis measurement of a backscattered electron-photon pair, given that the particles initially have opposite helicity and circular polarisation eigenvalues. %In fact, around the Compton IR divergence our numerical results seem to indicate a diverging QFI, and thus in principle the possibility to reach arbitrarily high precision.
\end{itemize}

\hspace*{2cm}

\FloatBarrier
\section{\label{sec:Conclusion} Conclusion and outlook}
We considered asymptotically-free fermions and photons that interact at the tree level, scatter, and then become asymptotically-free again before being measured. We then considered the quantum states of the scattered fermions and photons in, respectively, the helicity-basis and polarisation-basis Hilbert spaces, and computed the QFIM of these states with respect to the external parameters of the scattering (namely, the particles' momentum magnitude and the polar scattering angle), which are continuous. In this proof-of concept study we showed that the internal degrees of freedom of scattered fermions and photons carry information about their external degrees of freedom, and hence, placed precision limits on any unbiased estimation of the latter. We implicitly used quantum field operators and the formalism of QED in order to model the evolution of these particles states. However, the QFIM itself is a geometric property of the Hilbert space that the particle states are elements of. An interesting continuation of this investigation would be to consider the QFIM (on the particle-state space) with respect to the coupling constant of the quantum field itself. This would describe how deformations of the quantum field Lagrangian affect the geometry of the related particle-state Hilbert space, and would hence reveal a fundamental feature of the quantum field theory under consideration.

An alternative line of development would be to apply the method put forward in this paper to other quantum fields. Ref. \cite{nunez2025universality} showed that two gluons in a pure state are entangled in their polarisations as a result of three and four-gluon vertex interactions, given that they have opposite polarisations before the interaction. The scattered state derived in Ref. \cite{nunez2025universality} is a function parametrised by the polar scattering angle and the gluons' momenta in the centre-of-mass frame. This means that the QFIM of this state can be found using the method presented here. %However, in this case the QFIM would no longer be applicable to a physical measurement and parameter estimation scenario, as gluons are not asymptotically free. 

\acknowledgments

AS thanks M.G.~Genoni for the recurrent discussions and consultancies on the matter of quantum estimation. This work was funded by the Leverhulme Trust Research Project Grant RPG-2024-287.

\bibliography{apssamp}% Produces the bibliography via BibTeX.

\end{document}